\documentclass[twocolumn,secnumarabic,amssymb, nobibnotes, aps, prd]{revtex4-2}

\usepackage{amsmath}
\usepackage{amssymb}

\usepackage{graphicx}
\usepackage{dcolumn}
\usepackage{bm}
\usepackage[textsize=footnotesize]{todonotes}
\usepackage{tikz}
\usetikzlibrary{mindmap,shadows}
\usetikzlibrary{backgrounds}
\usetikzlibrary{arrows, calc}
\usetikzlibrary{shapes,shapes.geometric,shapes.misc}
\tikzstyle{none}=[inner sep=0mm]
\usepackage{drawmatrix}
\usepackage{centernot}

\usepackage{xr}

\tikzstyle{ph}=[fill=none, draw=none, shape=rectangle, text width=4 em]

\tikzstyle{Arrow}=[->]

\usepackage[nomargin,inline,marginclue,draft,multiuser]{fixme}
\fxsetup{theme=color, mode=multiuser} 
\FXRegisterAuthor{pb}{apb}{\color{red}PB} 
\FXRegisterAuthor{mb}{amb}{\color{blue}MB}
\FXRegisterAuthor{rf}{arf}{\color{brown}R1} 
\FXRegisterAuthor{rs}{ars}{\color{teal}R2}

\usepackage[normalem]{ulem}

\graphicspath{ {./}{../figures/}}

\begin{document}


\title{Directional coupling detection through cross-distance vectors}

\author{Martin Bre\v{s}ar}%
\email{martin.bresar@ijs.si}
\affiliation{ 
Jo\v{z}ef Stefan Institute,
Jamova cesta 39, SI-1000 Ljubljana, Slovenia
}%
\affiliation{ 
Jo\v{z}ef Stefan International Postgraduate School, Jamova cesta 39, SI-1000 Ljubljana, Slovenia
}
\author{Pavle Bo\v{s}koski}%
\affiliation{ 
Jo\v{z}ef Stefan Institute,
Jamova cesta 39, SI-1000 Ljubljana, Slovenia
}%


\begin{abstract}
Inferring the coupling direction from measured time series of complex systems is challenging.
We propose a new state space based causality measure obtained from \textit{cross-distance vectors} for quantifying interaction strength.
It is a model-free noise-robust approach that requires only a few parameters.
The approach is applicable to bivariate time series and is resilient to artefacts and missing values.
The result is two \textit{coupling indices} that quantify coupling strength in each direction more accurately than the already established state space measures.
We test the proposed method on different dynamical systems and analyse numerical stability.
As a result, a procedure for optimal parameter selection is proposed, circumventing the challenge of determining the optimal embedding parameters.
We show it is robust to noise and reliable in shorter time series. 
Moreover, we show that it can detect cardiorespiratory interaction in measured data.
A numerically efficient implementation is available at \url{https://repo.ijs.si/mbresar/cd-vec}.
\end{abstract}

\maketitle

\section{Introduction}

Complex systems found in nature can often be considered as many interacting subsystems.
Subsystems are often inherently connected and can not be considered isolated from each other, which raises the question of how they interact with each other.
In this article, we propose a new method that belongs to the family of state space distance approaches. 
It is capable of detecting and quantifying interactions in a computationally efficient way.
It can be applied to bivariate time series to quantify the coupling strength in both directions.
Furthermore, it is applicable to both linear and nonlinear coupling.

When considering two subsystems, there are four possibilities for the direction of their interaction.
They can be independent, unidirectionally coupled (in either direction), or bidirectionally coupled.
Another property often of interest is the nature of coupling, e.g., linear or nonlinear.
Additionally, the coupling can be time-dependent, which poses a new problem in detecting it.
The problem becomes even more complex in the case of more than two subsystems.


Typically, we measure a subsystem.
For example, consider the system of a human body.
We can characterise the cardiac subsystem by measuring the heart's electrical activity with an electrocardiogram.
By performing such measurements for each subsystem, we obtain time series.
The goal is to infer the direction and the nature of the interactions between the subsystems from measured time series.


This article shows that the proposed coupling indices can infer the coupling direction for various regular and chaotic systems.
It can quantify the coupling strength in both directions for unidirectional or bidirectional coupling.
It turns out to be noise robust and also efficient for shorter time series.
Most importantly, compared to the most prominent state space approaches, the proposed indices more accurately and reliably detect the coupling direction.
We discuss the possible pitfalls of state space approaches in detail.
We present an in-depth analysis of the behaviour of the proposed quantities.
As a result, we propose a procedure of selecting optimal parameter values, thus achieving a robust performance.
In such a way, we circumvent the problem of optimal embedding parameters selection.
Additionally, we present a means of dealing with artefacts or missing values.
Finally, we test the effectiveness of the approach on measured cardiorespiratory data.

\subsection*{Related work}

Identifying causal relationships arises in different fields that deal with complex systems.
For this purpose, different methods for detecting coupling between subsystems are being developed.
Some of the most widely used are Granger causality, information theory, phase dynamics, and state space methods.
Prominent examples of areas where they are applicable are physiology \cite{Runge2015, Schulz2013}, neuroscience \cite{Wibral2014}, earth system sciences \cite{Runge2019}, ecology, \cite{Sugihara2012, Ye2015} and economics \cite{Granger1969, Hoover2008}.

A brief overview of coupling detection methods is given below.
For a more comprehensive overview of the available approaches, the prospective reader is referred to the overviews \cite{Papana2021, Clemson2016} and references therein.

\paragraph*{Granger causality}
Coupling detection methods often follow the idea of Granger causality \cite{Granger1969}.
The method in the original work is based on fitting a vector autoregressive model.
Based on this idea, numerous other methods were proposed, such as Partial directed coherence \cite{Baccala2001}, which is a multivariate frequency approach to the original Granger causality.

\paragraph*{Information transfer}
A widely used family of methods for studying interactions is based on information theory.
Most commonly studied is the transfer entropy \cite{Schreiber2000}, which quantifies information transfer from one subsystem to another.
It is mathematically equivalent to conditional mutual information \cite{Palus2007}.
It is a model-free method that can detect both linear and nonlinear coupling.
Entropy methods rely on estimating multidimensional probability distributions, which can be challenging for shorter time series.

\paragraph*{Phase dynamics}
For oscillatory systems, methods based on the oscillation phase can be used.
\citet{Rosenblum2001} proposed a method for quantifying asymmetry in the interaction between two oscillating subsystems.
Phase transformation can be used together with information measures to more accurately quantify information flow between oscillators \cite{Palus2003}.
Time-frequency approaches can also be used, such as wavelet phase coherence \cite{Bandrivskyy2004} for detecting linear and wavelet bispectral analysis \cite{Jamsek2007} for detecting nonlinear interactions.

\paragraph*{State space methods}

In weakly coupled bivariate systems, close states of the driven subsystem are mapped to close states of the driving subsystem.
The opposite effect is much smaller.
Different methods quantifying this effect in both directions have been proposed \cite{Arnhold1999, Andrzejak2003, Chicharro2009}.

\paragraph*{Bivariate versus multivariate}
An important distinction is between bivariate and multivariate methods.
In the case of more than two interacting subsystems, all of them must be accounted for when analysing interactions.
This necessity led to the multivariate approach of many of the mentioned approaches.
Partial granger causality \cite{Guo2008} is an extension to the Granger causality that excludes the effects of latent variables.
Causation entropy \cite{Sun2014} is a measure similar to transfer entropy that considers multiple variables.
Multivariate methods, however, often lead to the curse of dimensionality, which means the estimation of multivariate measures becomes increasingly problematic with an increasing dimension of the measured system.


\section{Cross-distance vectors} \label{sectiontse}


Consider a pair of unidirectionally coupled subsystems $\bm{x}(t) = (x_1(t), \dots, x_{n_x}(t))$ and $\bm{y}(t) = (y_1(t), \dots, y_{n_y}(t))$.
If the coupling direction is $y \rightarrow x$, their time evolution is described by

\begin{subequations}\label{equation1}
     \begin{align}
      & \frac{d\bm{x}(t)}{dt} = \bm{f}(\bm{x},t) + \bm{g}(\bm{x},\bm{y}) \label{equation1a}\\
      &\frac{d\bm{y}(t)}{dt} = \bm{h}(\bm{y},t), \label{equation1b}
     \end{align}
    \end{subequations}

where $\bm{g}(\bm{x},\bm{y})$ is the coupling function.
We consider time-independent coupling functions, though they generally can depend on time.
Our goal is to define a measure for the magnitude and direction of coupling from the observed time series.
For $\bm{g}(\bm{x},\bm{y}) \neq 0$, the subsystems and, therefore, their trajectories are not independent. 


The trajectories of both subsystems $\bm{x}(t)$ and $\bm{y}(t)$ are observed at equally spaced times $t_i = t_0 + i\Delta t$ for $i\in \mathbb{Z}$. 
Thus, we obtain time series of these trajectories $(\bm{x}(t_i) : i \in \mathbb{Z})$ and  $(\bm{y}(t_i): i \in \mathbb{Z})$.
We assume that $\Delta t$ is sufficiently small to capture all the necessary information.
The influence of $\Delta t$ and potential downsampling are described in detail in Section \ref{sectioncompdetails}.
Furthermore, we assume that the signals do not contain any trends, no commensurate frequency components, and other trivial artefacts that can be easily removed by a simple preprocessing.
 


Most often, only one dimension of a subsystem is measured.
Thus, a single time series of length $N$ is obtained.
Therefore, we consider time series of one-dimensional values and omit the bold notation.

We split the time series into segments of length $L$.
There is a total of $N-L+1$ segments for each time series.
Segments at a time moment $t_i$ are defined as
\begin{align}
& \bm{o}^x_i = (x(t_i), x(t_{i+1}), \dots, x(t_{i+L-1}))\label{xsegment} \\
& \bm{o}^y_i = (y(t_i), y(t_{i+1}), \dots, y(t_{i+L-1})).
\label{ysegment}
\end{align}

Furthermore, let us assume a measure of the similarity of two segments, i.e., a distance between two vectors
\begin{equation}
d(\bm{o}^x_i, \bm{o}^x_j) = \|\bm{o}^x_i - \bm{o}^x_j \|.
\label{dmetric}
\end{equation}
We ask two questions:
\begin{enumerate}
    \item If two segments of the driven subsystem $\bm{o}^x_i$ and $\bm{o}^x_j$ at times $t_i$ and $t_j$ are similar, are the segments at those times of the driving subsystem $\bm{o}^y_i$ and $\bm{o}^y_j$ also similar?
    \item And inversely: If two segments of the driving subsystem $\bm{o}^y_i$ and $\bm{o}^y_j$ at times $t_i$ and $t_j$ are similar, are the segments at those times of the driven subsystem $\bm{o}^x_i$ and $\bm{o}^x_j$ also similar?
\end{enumerate}
To clarify, we ask whether the following statements are true in the coupling direction $y\rightarrow x$:
\begin{equation}
1. \:\:\:\:\:\:\:\: d(\bm{o}_i^x, \bm{o}_j^x) \approx 0 \implies d(\bm{o}_i^y, \bm{o}_j^y) \approx 0
\label{statement1}
\end{equation}
\begin{equation}
2. \:\:\:\:\:\:\:\: d(\bm{o}_i^y, \bm{o}_j^y) \approx 0 \implies d(\bm{o}_i^x, \bm{o}_j^x) \approx 0
\label{statement2}
\end{equation}

It is well known that statement \eqref{statement1} is {\bf true} and \eqref{statement2} is {\bf false} for weakly coupled oscillators \cite{Arnhold1999}.
We provide an explanation of why this is the case for different subsystems in Appendix~\ref{appendix}.
State space methods take advantage of this property to infer the coupling direction from measured time series.
In what follows, we propose a new measure that quantifies coupling strength in each direction more accurately than the already established state space indices.

\subsection{Cross-distance vectors algorithm}
First, we construct segments \eqref{xsegment} and \eqref{ysegment} and choose a distance measure \eqref{dmetric}.
In this article, we use
\begin{equation}
d(\bm{o}_i^x, \bm{o}_j^x) =\sqrt{ \frac{1}{L}\sum_{m=1}^{L} \big(x(t_{i+m-1}) - x(t_{j+m-1})\big)^2}.
\label{distanceequation}
\end{equation}
Next, we construct the distance matrix
\begin{equation}
\bm{D}^x, \: D^x_{ij} = d(\bm{o}^x_i, \bm{o}^x_j)
\label{eq:dist_x}
\end{equation}
of size $(N-L+1) \times (N-L+1)$, which contains the distances between all pairs of segments.
We similarly calculate the distance matrix of the other time series
\begin{equation}
\bm{D}^y, \: D^y_{ij} = d(\bm{o}^y_i, \bm{o}^y_j).
\label{eq:dist_y}
\end{equation}

The goal is to rearrange the $i$-th row of $\bm{D}^y$ in the same order as if the $i$-th row of $\bm{D}^x$ was to be sorted in ascending order.
This is achieved by calculating an index permutation $P^x_i$ so that the $i$-th row is sorted in ascending order under that permutation.
We do this for every row and obtain $N-L+1$ permutations.

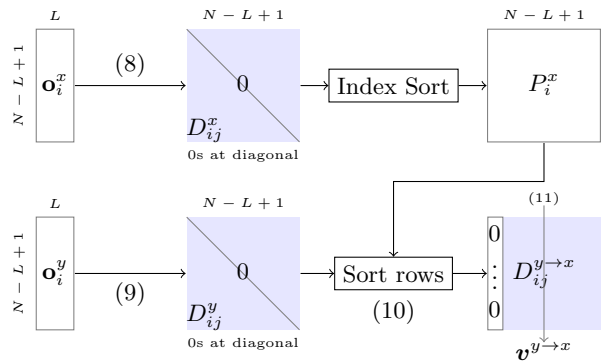
\begin{figure}
\begin{tikzpicture}
		\node [style=none] (9) {
		$
		\drawmatrix[width=.5, height=1.5] {\mathbf{o}_i^x}
		$
		};
		\node [style=none, right of=9, xshift=1.5cm] (10) {
		$
		\drawmatrix[diag, bbox/.append style={fill=blue!10}, width=1.5, height=1.5]0
		$
		};
		
		\node [style=none, anchor=south west] at (10.south west) {$D_{ij}^x$};
		
		\node [rectangle, draw=black,, right of=10, xshift=1cm] (id_sort) {Index Sort};
		\node [style=none, right of=id_sort, xshift=1cm] (pxi) {
		$
		\drawmatrix[width=1.5, height=1.5] {P_i^x}
		$
		};
		
		\node [style=none, below of=9, yshift=-1.5cm] (9y) {
		$
		\drawmatrix[width=.5, height=1.5] {\mathbf{o}_i^y}
		$
		};
		\node [style=none, below of=10, yshift=-1.5cm] (10y) {
		$
		\drawmatrix[diag, bbox/.append style={fill=blue!10}, width=1.5, height=1.5]0
		$
		};
		\node [style=none, anchor=south west] at (10y.south west) {$D_{ij}^y$};
		\node [rectangle, draw=black, below of=id_sort, yshift=-1.5cm] (id_sort_y) {Sort rows};
		\node [style=none, below of=id_sort_y, yshift=14] {\eqref{Dyvxeq}};
		\node [style=none, right of=id_sort_y, xshift=1cm] (final) {
		$
		\drawmatrixset{bbox style={fill=blue!10}}
		\drawmatrix[width=.2, height=1.5,bbox width=1.5,,bbox height=1.5]{D_{ij}^{y\to x}}
$
		};
		
		\node[style=none, above of=final] (st) {\tiny{\eqref{couplingvector}}};
		\node[style=none, below of=final, yshift=-1] (en) {
		$\bm{v}^{y\to x}$
		};
		
		\node[style=none, anchor=west, xshift=.5] at (final.west) {
		$
		\begin{matrix}
		0 \\ 
		\vdots \\
		0 \\
		\end{matrix}
		$
		};
		
		\node [style=none, below of=10y, yshift=2] {\tiny{0s at diagonal}};
		\node [style=none, below of=10, yshift=2] {\tiny{0s at diagonal}};
		
		\node [style=none, above of=9, yshift=-2] {\tiny $L$};
		\node [style=none, left of=9, rotate=90, yshift=-15] {\tiny $N-L+1$};
		\node [style=none, above of=10, yshift=-2] {\tiny $N-L+1$};

		\node [style=none, above of=pxi, yshift=-2] {\tiny $N-L+1$};
		
		\node [style=none, above of=9y, yshift=-2] {\tiny $L$};
		\node [style=none, left of=9y, rotate=90, yshift=-15] {\tiny $N-L+1$};
		\node [style=none, above of=10y, yshift=-2] {\tiny $N-L+1$};
		
		\path[->] (9) edge node[above]  {\eqref{eq:dist_x}} (10);
		\path[->] (9y) edge node[below]  {\eqref{eq:dist_y}} (10y);
		\path[->] (10) edge (id_sort);
		\path[->] (10y) edge (id_sort_y);
		\path[->] (id_sort) edge (pxi);
		\path[->] (id_sort_y) edge (final);
		
		\draw[style=Arrow,color=black!50] (st) -- (en);
		
		\draw[style=Arrow] (pxi.south) |- ($ (id_sort_y.north) + (0,1) $) -- (id_sort_y.north);
		
\end{tikzpicture}
\caption{Schematic representation of the algorithm for calculating the cross-distance vector $\bm{v}^{y\to x}$ in~\eqref{couplingvector}.}
\label{fig:algorithm}
\end{figure}

Now, we sort $i$-th row of $\bm{D}^y$ according to the permutation $P^x_i$ for all rows and obtain the matrix
\begin{equation}
\bm{D}^{y\rightarrow x}, \: D^{y\rightarrow x}_{ij} = D_{i,P_i^x(j)}^y\:\:.
\label{Dyvxeq}
\end{equation}
This matrix highlights whether segments of $y$ are similar at times at which segments of $x$ are similar.
For example, if the first few elements of $i$-th row of matrix \eqref{Dyvxeq} are the smallest entries in that row, this indicates that statement \eqref{statement1} is true.
To evaluate this for all rows, we finally average out the rows of the matrix to obtain the \textit{cross-distance vector}
\begin{equation}
\bm{v}^{y\rightarrow x}, \: v^{y\rightarrow x}_{j} = \frac{1}{N-L+1} \sum_i D^{y\rightarrow x}_{ij}.
\label{couplingvector}
\end{equation}
Similarly, we calculate the matrix $\bm{D}^{x\rightarrow y}$ and the vector $\bm{v}^{x\rightarrow y}$ by inverting the roles of $x$ and $y$.
The vector $\bm{v}^{y\rightarrow x}$ can be used to assess the truth of statement \eqref{statement1} and thus detect coupling in the direction $y\rightarrow x$.
Conversely, the vector $\bm{v}^{x\rightarrow y}$ can be used to assess the truth of statement \eqref{statement2} and thus detect coupling in the direction $x\rightarrow y$.
The complete algorithm is schematically shown in \figurename~\ref{fig:algorithm}.

It should be noted that the first entry $v_{0}^{y\rightarrow x}$ of cross-distance vectors~\eqref{couplingvector} will always be precisely zero due to zero diagonals of the distance matrices.
So, $v_{1}^{y\rightarrow x}$ is the first non-zero entry.
In the subsequent analysis, we will always omit the zero values in graphs.

\subsection{Coupling index}
\label{couplingindexsection}

\begin{figure}
\includegraphics[width = \linewidth]{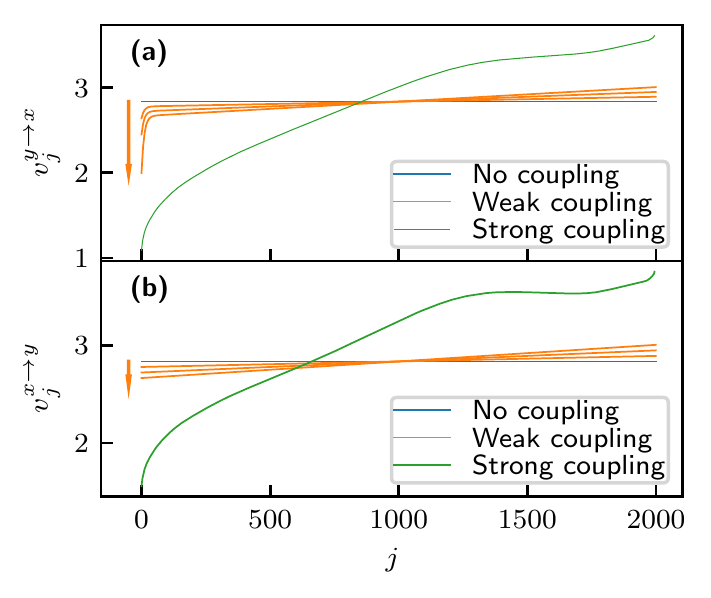}
\caption{A generic example of the behaviour of cross-distance vectors $\bm{v}^{y\rightarrow x}$ and $\bm{v}^{x\rightarrow y}$ in different limits of the coupling strength. The direction of coupling in the example is $y \rightarrow x$. The orange arrows represent increasing coupling strength for the three weak coupling examples. Note the appearance of tails in the beginning of $\bm{v}^{y\rightarrow x}$, but not in $\bm{v}^{x\rightarrow y}$.}
\label{figcouplingvectors}
\end{figure}

We can define two {\it coupling indices} to quantify coupling strength in each direction.
Ideally, an index is zero when there is no coupling and increases with increased coupling.
Let us again consider unidirectionally coupled subsystems \eqref{equation1}.
Mind that the coupling direction is $y \rightarrow x$, i.e., $x$ is the driven and $y$ is the driving subsystem.
Consider three limits of the coupling strength: no coupling, weak coupling and strong coupling.
The general behaviour of cross-distance vectors in these limits is shown in \figurename~\ref{figcouplingvectors}.

The two subsystems $\bm{x}$ and $\bm{y}$ are independent if there is no coupling, i.e., $\bm{g}(\bm{x}, \bm{y}) = 0$ in \eqref{equation1a}.
Therefore, one subsystem's sorting permutation $P_i^x$ is random for the subsystem $y$ and the cross-distance vector \eqref{couplingvector} is expected to be roughly constant.
In the limit of infinite time series, both $\bm{v}^{y\rightarrow x}$ and $\bm{v}^{x\rightarrow y}$ will be constant.
There are examples where this is not the case (e.g. uncoupled subsystems which contain an oscillatory component with the same frequency), but these are exceptions.

In the limit of strong coupling, the term $\bm{g}(\bm{x}, \bm{y})$ in \eqref{equation1a} causes the subsystems to synchronise.
In this case, the cross-distance vectors generally cannot be used to infer the coupling direction.

If there is \textbf{weak coupling} in the direction $y \rightarrow x$, the values of $\bm{v}^{y\rightarrow x}$ change significantly from a constant since statement \eqref{statement1} is true.
The beginning of $\bm{v}^{y\rightarrow x}$ decreases (note the initial tails in \figurename~\ref{figcouplingvectors} (a)).
Apart from this initial tail, the bulk of $\bm{v}^{y\rightarrow x}$ remains roughly constant (though often gains an increasing trend).
Conversely, the other cross-distance vector $\bm{v}^{x\rightarrow y}$ stays roughly constant without the initial tail since statement \eqref{statement2} is false.
This is shown in \figurename~\ref{figcouplingvectors} (b).

The \textbf{decrease in the initial values of} of $\bm{v}^{y\rightarrow x}$ is what allows us to detect the coupling direction from the cross-distance vectors,
since the tail is present due to the term $\bm{g}(\bm{x}, \bm{y}) \neq 0$ in equation \eqref{equation1a}.
Mind that if a similar term also exists in \eqref{equation1b} (which means that coupling is bidirectional), the initial values of both cross-distance vectors $\bm{v}^{y\rightarrow x}$ and $\bm{v}^{x\rightarrow y}$ exhibit an initial tail.
To obtain the coupling indices, i.e., two values quantifying the detected coupling strength in each direction, we must somehow quantify this effect.
While this can be done in different ways, we propose a simple but effective approach.

We define the coupling index $c^{y\rightarrow x}$, which quantifies coupling strength in the direction $y\rightarrow x$, as the normalised difference between the means of $v^{y\rightarrow x}_{j}$ for small $j$ and for larger $j$
\begin{equation}
c^{y\rightarrow x} = \frac{V^{y\rightarrow x}_{k_1+1:k_2} - V^{y\rightarrow x}_{1:k_1}}{V^{y\rightarrow x}_{1:k_2}},
\label{Dequation}
\end{equation}
where
\begin{equation}
V^{y\rightarrow x}_{i:j} = \frac{1}{j-i+1}\sum_{k=i}^{j}v_k^{y\rightarrow x}.
\end{equation}
The possible range of $k_1$ and $k_2$ is $1 \leq k_1 < k_2 \leq N-L$.

To quantify coupling strength in the other direction, we calculate $c^{x\rightarrow y}$ by swapping $x$ and $y$.
Defined in this way, coupling indices are \textbf{zero in the absence of coupling} and \textbf{increase with increasing coupling strength} (at least in the limit of weak coupling).
Due to normalisation, indices will roughly range from $0$ to $1$.
They can be slightly negative if there is no coupling due to finite time series, as seen in most examples in this article.
This is, however, not a problem, as a negative index can be taken as a strong indicator of the absence of coupling.

The purpose of equation \eqref{Dequation} is to quantify the prominence of the initial tail.
The parameter $k_1$ defines the part of $\bm{v}^{y\rightarrow x}$ that includes only the initial tail. 
Conversely, $k_2$ defines the subsequent part, where the rate of increase of $\bm{v}^{y\rightarrow x}$ is significantly smaller.

One might argue that this selection is arbitrary.
Other options, such as kurtosis, could be explored to quantify the prominence of the initial tail.
For simplicity, we quantify coupling strength with \eqref{Dequation} in the subsequent analysis.

\paragraph*{Related state space causality measure}

From the family of the state space approaches, our index \eqref{Dequation} is closest to $M(Y|X)$ \cite{Andrzejak2003}.
It will therefore be compared to and used as a benchmark to test the reliability of the proposed indices $c$.
Using our notation, $M(Y|X)$ is defined as 
\begin{equation}
M(Y|X) = \frac{1}{N-L+1} \sum_{i=1}^{N-L+1} \frac{R_i(Y) - R_i^{k}(Y|X)}{R_i(Y) - R_i^k(Y)},
\label{Mequation}
\end{equation}
where
\begin{subequations}
\label{Mparts}
\begin{align}
& R_i(Y) = \frac{1}{N-L}\sum_{j=1,j\neq i}^{N-L+1} D_{ij}^y, \label{eqRi} \\
& R_i^k(Y) = \frac{1}{k}\sum_{j=2}^{k+1} D_{ij}^{y\rightarrow y}, \\
& R_i^k(Y|X) = \frac{1}{k}\sum_{j=2}^{k+1} D_{ij}^{y\rightarrow x}.
\end{align}
\end{subequations}
Roughly speaking, $M(Y|X)$ quantifies the mean of the initial tail of $\bm{v}^{y\rightarrow x}$ compared to the mean of the whole vector, which is highly influenced by possible trends in the vector.
This undesired property is also present in other state space measures.
On the other hand, the indices \eqref{Dequation} quantify only the prominence of the initial tail, \textbf{ignoring possible trends} and thus resulting in smaller values in the direction without coupling, which is desired.


\subsection{An example}\label{ssex}

Consider the system of analytically solvable unidirectionally coupled harmonic oscillators.
\begin{align}
& \ddot{x} = -\omega_1^2 x + \epsilon(y-x) \\
& \ddot{y} = -\omega_2^2 y
\end{align}
The coupling direction is $y\rightarrow x$.
The solution to these equations are oscillations with frequencies $\omega_2$ and $\sqrt{\omega_1^2+\epsilon}$.
For example, given the initial conditions $x(0) = 1, \dot{x}(0) = 0, y(0) = 1, \dot{y}(0) = 0$, the solution is
\begin{equation}
\begin{split}
x(t) & = \frac{1}{\epsilon + \omega_1^2 - \omega_2^2} \big(\epsilon \cos(\omega_2 t) + (\omega_1^2-\omega_2^2) \cos(\sqrt{\epsilon+\omega_1^2} t)\big) \\
y(t) & = \cos(\omega_2 t).
\end{split}
\label{HO2}
\end{equation} 
Since we have the analytical solution, we can verify statements \eqref{statement1} and \eqref{statement2} directly.

Two segments of $x$ will be similar when the phases of both its oscillatory components will match, one of them being $\cos(\omega_2 t)$.
Since this is also the component of $y(t)$, its segments at those times will also be similar.
This confirms that the statement \eqref{statement1} is indeed \textbf{true}.

If two segments of $y(t)$ are similar, the phase of only one oscillatory component of $x(t)$ will match.
In contrast, the phase of the other component can take any value (unless the frequencies of the components of $x(t)$ are commensurable, which is an exception).
This confirms that the statement \eqref{statement2} is indeed \textbf{false}.

\begin{figure}
\includegraphics[width = \linewidth]{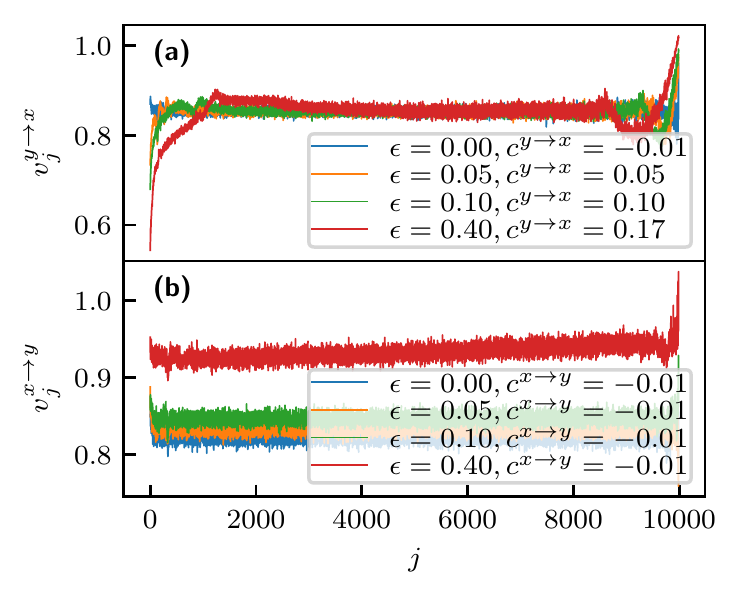}
\caption{The cross-distance vectors for the system of two coupled harmonic oscillators \eqref{HO2}. The system parameters are $\omega_1 = 0.83, \omega_2 = 2.11$, the time series parameters are $N = 10^4,  \Delta t = 0.05$, the segment length is $L = 10$, and the coupling index parameters are $k_1 = 100, k_2 = 1000$.}
\label{hocv}
\end{figure}

Numerically calculated cross-distance vectors are shown in \figurename\ref{hocv}.
As expected, the beginning of $\bm{v}^{y\rightarrow x}$ is significantly smaller than the bulk (\figurename\ref{hocv}~(a)), and $\bm{v}^{x\rightarrow y}$ stays roughly constant at nonzero coupling (\figurename\ref{hocv}~(b)).

Additionally, \figurename\ref{hocv} shows that the coupling index of the direction of coupling $c^{y\rightarrow x}$ increases with increased coupling parameter $\epsilon$.
In contrast, the coupling index in the direction without coupling $c^{x\rightarrow y}$ remains roughly constant.
By simply comparing the values, we can correctly determine the coupling direction.

We also notice that the end of $\bm{v}^{y\rightarrow x}$ has larger values than the bulk of the vector, which indicates that a statement similar to \eqref{statement1} is also true for dissimilarity: if two segments at times $t_i$ and $t_j$ of the driven subsystem are very dissimilar, the segments at these times of the driving subsystem are more likely also to be dissimilar.
This opposite effect is, however, not as expressed and does not appear in all systems.

When numerically calculating the cross-distance vectors, we added Gaussian noise $\mathcal{N}(0,\sigma^2=10^{-4})$ to the time series.
In noiseless periodic systems, cross-distance vectors can exhibit periodic oscillations.
By adding small noise, they disappear, and the vectors behave the same as in chaotic systems.
We still used this example due to the availability of the analytical solution.
In practice, periodic systems are of lesser interest and always contain noise.

In the following section, we will present a numerical analysis that shows that cross-distance vectors can be used to determine the direction of coupling in different systems.

\section{Different systems analysis} \label{secnuman}

The previous section shows that the cross-distance vectors can detect the coupling direction in the coupled harmonic oscillators system.
Here, we will show that detection is possible not only for regular subsystems but also for discrete chaotic subsystems and for autonomous and nonautonomous continuous chaotic subsystems.
For an overview on chaotic dynamical systems, the reader is referred to \cite{Strogatz2015, Ott2002}.
We chose systems that have already been analysed with a coupling detection method.
The indices $M(Y|X)$ and $M(X|Y)$ will be shown next to $c^{y\rightarrow x}$ and $c^{x\rightarrow y}$.
At the end of this section, these two methods will be compared.

\subsection{Hénon maps}
\begin{figure}
\includegraphics[width = \linewidth]{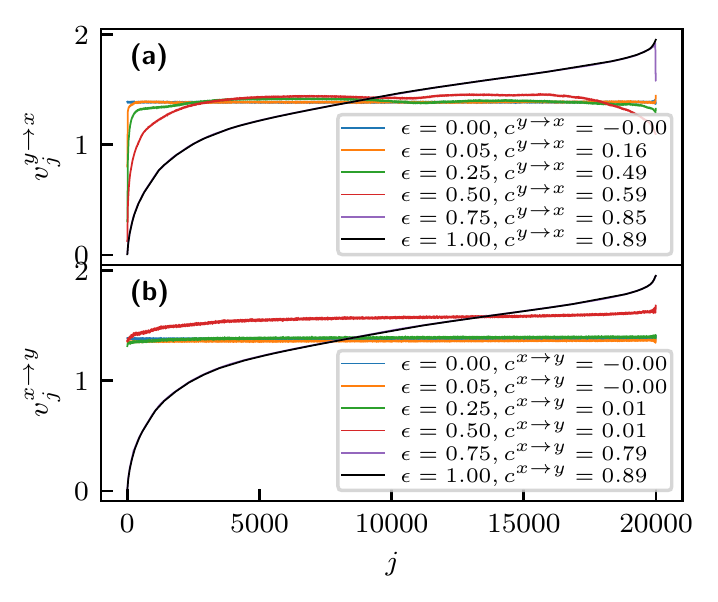}
\caption{The cross-distance vectors for unidirectionally coupled Hénon maps at different values of the coupling parameter $\epsilon$. They are obtained from the time series of the subsystems' coordinates $x_1$ and $y_1$ and with segment length $L = 10$.}
\label{henonvectors}
\end{figure}

Hénon map is a discrete-time dynamical system.
Based on the values of its two parameters $a$ and $b$ it can be chaotic or converge to a periodic orbit.
We choose the most commonly studied map with parameter values $a = 1.4$ and $b = 0.3$, which yield chaotic dynamics.
Two unidirectionally coupled maps are defined by four equations.
The driven subsystem is described by
\begin{align}
& x_{1}' = a - \big(\epsilon x_1 y_1 + (1 - \epsilon) x_1^2 \big) + b x_2 \\
& x_2' = x_1,
\end{align}
and the driving one by
\begin{align}
& y_{1}' = a - y_1^2 + b y_2 \\
& y_2' = y_1
\end{align}
The coupling direction is $y\rightarrow x$.
This system was analysed in \cite{Palus2001}.
We choose time series length $N = 2\cdot 10^4$ and segment length $L = 10$.
The cross-distance vectors at a few values of the coupling parameter $\epsilon$ are shown in \figurename\ref{henonvectors}.
They behave as expected.
In the absence of coupling, they are both constant.
For weak coupling, the initial values of $\bm{v}^{y\rightarrow x}$ decrease with increased coupling, while this does not happen with $\bm{v}^{x\rightarrow y}$.
For strong coupling, both $\bm{v}^{y\rightarrow x}$ and $\bm{v}^{x\rightarrow y}$ lose the straight shape, and the coupling direction cannot be inferred from them.

\begin{figure}
\includegraphics[width = \linewidth]{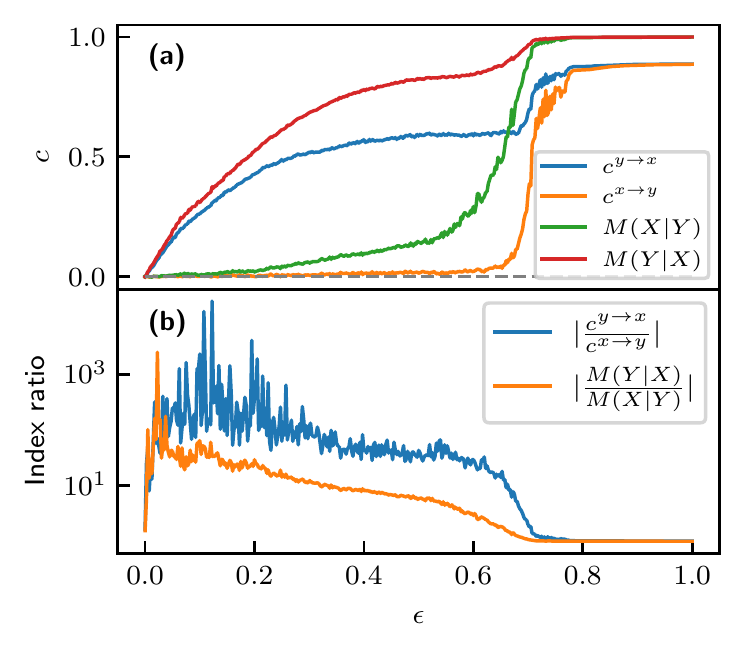}
\caption{The indices $c^{y\rightarrow x}$, $c^{x\rightarrow y}$, $M(Y|X)$, and $M(X|Y)$ (a), and index ratios (b) for unidirectionally coupled Hénon maps at different values of the coupling parameter $\epsilon$.
The coupling indices $c$ were calculated by \eqref{Dequation} with $k_1 = 10, k_2 = 100$, and the indices $M$ were calculated by \eqref{Mequation} with $k=10$.
}
\label{henonD}
\end{figure}

The dependence of the coupling indices $c^{x\rightarrow y}$ and $c^{y\rightarrow x}$ on the coupling parameter $\epsilon$ is shown in \figurename \ref{henonD} (a).
In the absence of coupling, both coupling indices are close to zero.
With increased coupling, $c^{y\rightarrow x}$ increases for small values of $\epsilon$.
The other coupling index $c^{x\rightarrow y}$ is close to zero until the synchronisation threshold at $\epsilon \approx 0.75$.
Therefore, the cross-distance vectors can be used to infer the coupling direction in unsynchronised Hénon maps correctly.

\subsection{Rössler systems}
Rössler system is a system of three nonlinear ordinary differential equations, which define a chaotic continuous-time dynamical system.
Two coupled Rössler subsystems are defined by six equations.
We chose the same parameters for the subsystems as in \cite{Palus2007}.
The driven subsystem is thus described by
\begin{align}
& \dot{x}_{1} = - 0.985 x_{2} - x_{3} + \epsilon(y_{1} - x_{1}) \\
& \dot{x}_{2} = 0.985 x_{1} + 0.15 x_{2} \\
& \dot{x}_{3} = 0.2 + x_{3}(x_{1} - 10).
\end{align}
and the driving subsystem by
\begin{align}
& \dot{y}_{1} = - 1.015 y_{2} - y_{3} \\
& \dot{y}_{2} = 1.015 y_{1} + 0.15 y_{2} \\
& \dot{y}_{3} = 0.2 + y_{3}(y_{1} - 10).\end{align}

Coupling is unidirectional in the direction $y\rightarrow x$.
It should be noted that the first subsystem is in a regular regime for $\epsilon=0$, and in a chaotic regime for $\epsilon>0$, while the second subsystem is always chaotic.
Therefore, we omit analysis at zero coupling.

Time series parameters are $N = 2\cdot 10^4, \Delta t = 0.5$.
The system integration was done with the Runge-Kutta 4 integrator with time step 0.01.
We choose the time series of the subsystems' coordinates $x_1$ and $y_1$ and segment length $L = 20$ for the calculation of the cross-distance vectors.

\begin{figure}
\includegraphics[width = \linewidth]{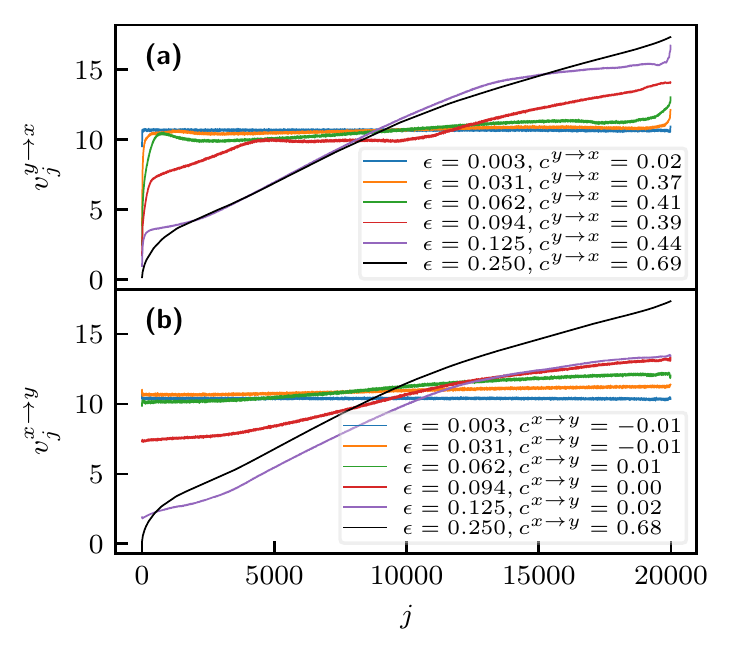}
\caption{The cross-distance vectors $\bm{v}^{y\rightarrow x}$ and $\bm{v}^{x\rightarrow y}$ for unidirectionally coupled Rössler subsystems at different values of the coupling parameter $\epsilon$.
They are obtained from the time series of the subsystems' coordinates $x_1$ and $y_1$ and with segment length $L = 20$.}
\label{rosslervectors}
\end{figure}

The behaviour of the cross-distance vectors for a few values of the coupling parameter $\epsilon$ is shown in \figurename\ref{rosslervectors}.
It is similar to the examples seen so far.
For weak coupling, the initial values of $\bm{v}^{y\rightarrow x}$ decrease from the bulk, while $\bm{v}^{x\rightarrow y}$ stays roughly constant.
For strong coupling, both subsystems become identical, and so do both cross-distance vectors.

\begin{figure}
\includegraphics[width = \linewidth]{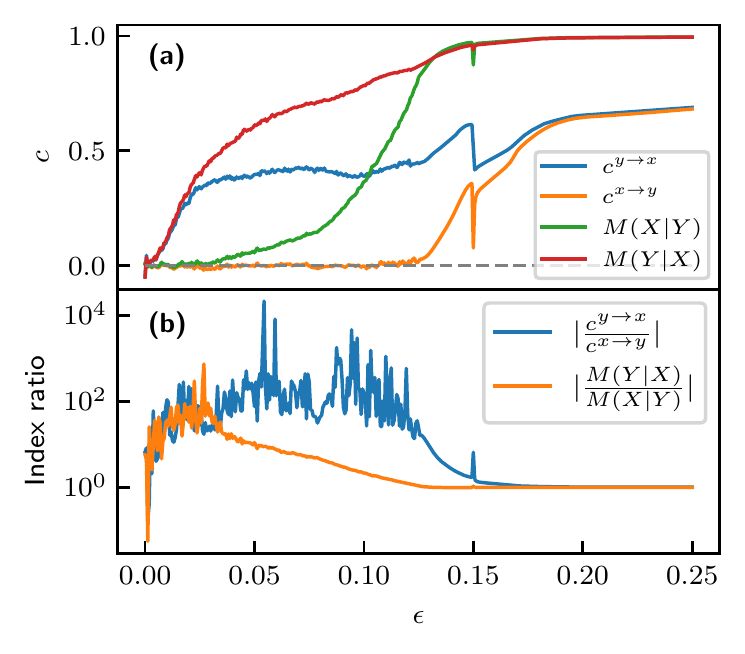}
\caption{The indices $c^{y\rightarrow x}$, $c^{x\rightarrow y}$, $M(Y|X)$, and $M(X|Y)$ (a), and index ratios (b) for unidirectionally coupled Rössler subsystems at different values of the coupling parameter $\epsilon$.
The coupling indices $c$ were calculated by \eqref{Dequation} with $k_1 = 10, k_2 = 100$, and the indices $M$ were calculated by \eqref{Mequation} with $k=10$.
}
\label{rosslerD}
\end{figure}

The dependence of the coupling indices $c^{x\rightarrow y}$ and $c^{y\rightarrow x}$ on the coupling parameter $\epsilon$ is shown in \figurename \ref{rosslerD} (a).
Similar to the previous example, $c^{y\rightarrow x}$ increases with increased coupling strength while $c^{x\rightarrow y}$ stays close to zero.
At $\epsilon \approx 0.15$, synchronisation occurs, and the coupling direction can no longer be inferred from the coupling indices.

This analysis shows that the cross-distance vectors can again be used to infer the coupling direction in unsynchronised Rössler subsystems correctly.

\subsection{Duffing systems} \label{ssduffing}
Duffing system is a periodically forced nonlinear oscillator with damping.
It is a nonautonomous continuous-time dynamical system.
It can exhibit chaotic or periodic dynamics based on the values of its parameters.
We chose the parameters of the two subsystems the same as in \cite{nasclanek}, resulting in coupled chaotic subsystems.
They are described by
\begin{align}
\label{eqcoupduff1}
&\ddot{x} + 0.2 \dot{x} - x + x^3 = 0.3 \cos(t) + \epsilon_1(y - x) \\
&\ddot{y} + 0.3 \dot{y} - y + y^3 = 0.5 \cos(1.2 t) + \epsilon_2(x - y).
\label{eqcoupduff2}
\end{align}
We consider a unidirectional case with $\epsilon_2 = 0$ and a bidirectional case with $\epsilon_2 = 0.1$.

Time series parameters are $N = 5\cdot 10^4$ and $\Delta t = 0.5$.
The system integration was done with the Runge-Kutta 4 integrator with a time step of $0.01$.
The time series of the first coordinates (positions) and segment length $L = 20$ were chosen for calculating the cross-distance vectors.
Since the behaviour of the cross-distance vectors is very similar to the above examples, here we only consider the coupling index dependences.

\begin{figure}
\includegraphics[width = \linewidth]{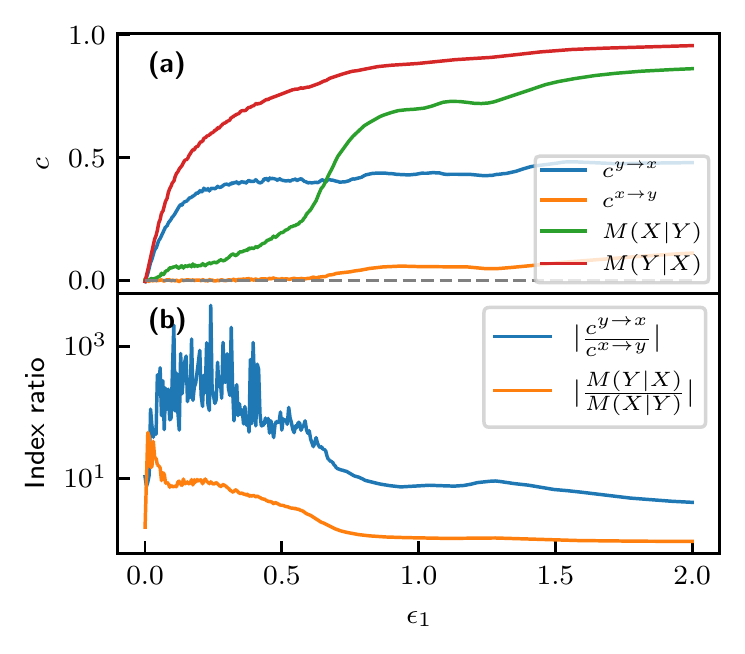}
\caption{The indices $c^{y\rightarrow x}$, $c^{x\rightarrow y}$, $M(Y|X)$, and $M(X|Y)$ (a), and index ratios (b) for unidirectionally coupled Duffing subsystems at different values of the coupling parameter $\epsilon_1$.
The coupling indices $c$ were calculated by \eqref{Dequation} with $k_1 = 10, k_2 = 100$, and the indices $M$ were calculated by \eqref{Mequation} with $k=10$.
}
\label{duffing1D}
\end{figure}

\begin{figure}
\includegraphics[width = \linewidth]{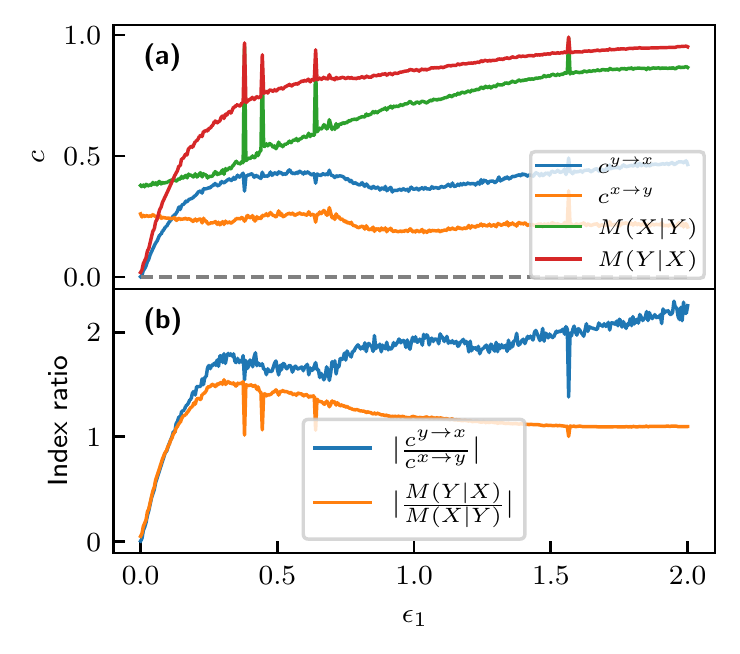}
\caption{The indices $c^{y\rightarrow x}$, $c^{x\rightarrow y}$, $M(Y|X)$, and $M(X|Y)$ (a), and index ratios (b) for bidirectionally coupled Duffing subsystems at different values of the coupling parameter $\epsilon_1$ and at $\epsilon_2 = 0.1$.
The sudden increases in all the indices, seen in four values of $\epsilon_1$, are due to system bifurcations.
The coupling indices $c$ were calculated by \eqref{Dequation} with $k_1 = 10, k_2 = 100$, and the indices $M$ were calculated by \eqref{Mequation} with $k=10$.
}
\label{duffing2D}
\end{figure}

The dependence of the coupling indices $c^{x\rightarrow y}$ and $c^{y\rightarrow x}$ on the coupling parameter $\epsilon_1$ is shown in \figurename \ref{duffing1D} (a) (unidirectional case) and \figurename \ref{duffing2D} (a) (bidirectional case).
In the unidirectional case, $c^{y\rightarrow x}$ again increases with increased coupling strength while $c^{x\rightarrow y}$ stays close to zero.
Synchronisation occurs at around $\epsilon_1 \approx 0.75$.
Therefore, cross-distance vectors can also be used to infer the coupling direction in nonautonomous Duffing subsystems correctly.

In the bidirectional case, $c^{x\rightarrow y}$ always has a positive value, while $c^{y\rightarrow x}$ increases with increased $\epsilon_1$ similar to the unidirectional case.
At around $\epsilon_1 = \epsilon_2 = 0.1$, the values of the coupling indices are the same.
Therefore, cross-distance vectors can also be used to quantify coupling strength in each direction in bidirectionally coupled subsystems.

\subsection{Comparison to established indices}
\label{seccvsm}

We have discussed the behaviour of the dependences of the coupling indices on the coupling parameter for various systems.
The comparison between the coupling indices $c$ and the established indices $M$ is shown in \figurename \ref{henonD} (a), \figurename \ref{rosslerD} (a), \figurename \ref{duffing1D} (a) and \figurename \ref{duffing2D} (a).
Both index variants were obtained from the same segments.
The choice of the parameter $k$ in \eqref{Mequation} was done in a similar manner as the choice of $k_1$ in \eqref{Dequation}.
It turns out that the optimal value of $k$ is similar to the optimal value of $k_1$.
This is not surprising since $k_1$ is chosen such that it captures the initial tail, much like $k$.
In all of the examples in this article, $k_1 = k = 10$.

In the unidirectional cases, both indices behave similarly in the sense that in the direction $y\rightarrow x$, they increase with increased coupling, while the index in the other direction stays close to zero.

An important difference is between $c^{x\rightarrow y}$ and $M(X|Y)$.
Ideally, these indices should be zero when the coupling direction is $y\rightarrow x$.
The new index $c^{x\rightarrow y}$ is significantly smaller than $M(X|Y)$.
Most importantly, the ratio $c^{y\rightarrow x} / c^{x\rightarrow y}$ is significantly larger than the ratio $M(Y|X) / M(X|Y)$, which is shown in \figurename \ref{henonD} (b), \figurename \ref{rosslerD} (b), and \figurename \ref{duffing1D} (b).
As explained in Section~\ref{couplingindexsection}, the summation in \eqref{eqRi} goes up to $N-L+1$, which captures the increasing trend seen in the cross-distance vectors and leads to larger values of $M(X|Y)$.

Ideally, the index ratios should be infinite.
By comparing the coupling indices, the coupling direction is more accurately determined with the $c$ indices.


In the bidirectional case, the main difference between the two methods is in the indices of the direction $x\rightarrow y$.
Ideally, they should be constant since $\epsilon_2 = 0.1$ is constant.
The coupling parameter $c^{x\rightarrow y}$ does vary slightly, but less than $M(X|Y)$.
Interestingly, bifurcations have a small impact on the coupling indices $c$ and a large impact on $M$.

\section{Numerical stability}
\label{secnumstab}

We have shown that the cross-distance vectors can detect the coupling direction from measured time series.
In this section, we will discuss the numerical stability and parameter selection of this method.
All the analysis will be done on the \textbf{test system} of unidirectionally coupled Duffing subsystems with the coupling parameters $\epsilon_1 = 0.1$ and $\epsilon_2 = 0$, where the system integration is done with the Runge-Kutta 4 integrator and then sampled at $\Delta t = 0.5$.
As in the previous section, the position coordinates' time series will be used to obtain the cross-distance vectors.
We chose this system because it consists of continuous chaotic subsystems, a property commonly found in real systems. 
However, analysis of all systems in Section \ref{secnuman} gives similar results.
The analysis in this section focuses solely on the cross-distance vectors and the new index $c$.
A detailed comparison of numerical properties of the indices $c$ and $M$ is done in Appendix \ref{appendixB}.

\subsection{Time series length}
\label{secNdependence}

\begin{figure}
\includegraphics[width = \linewidth]{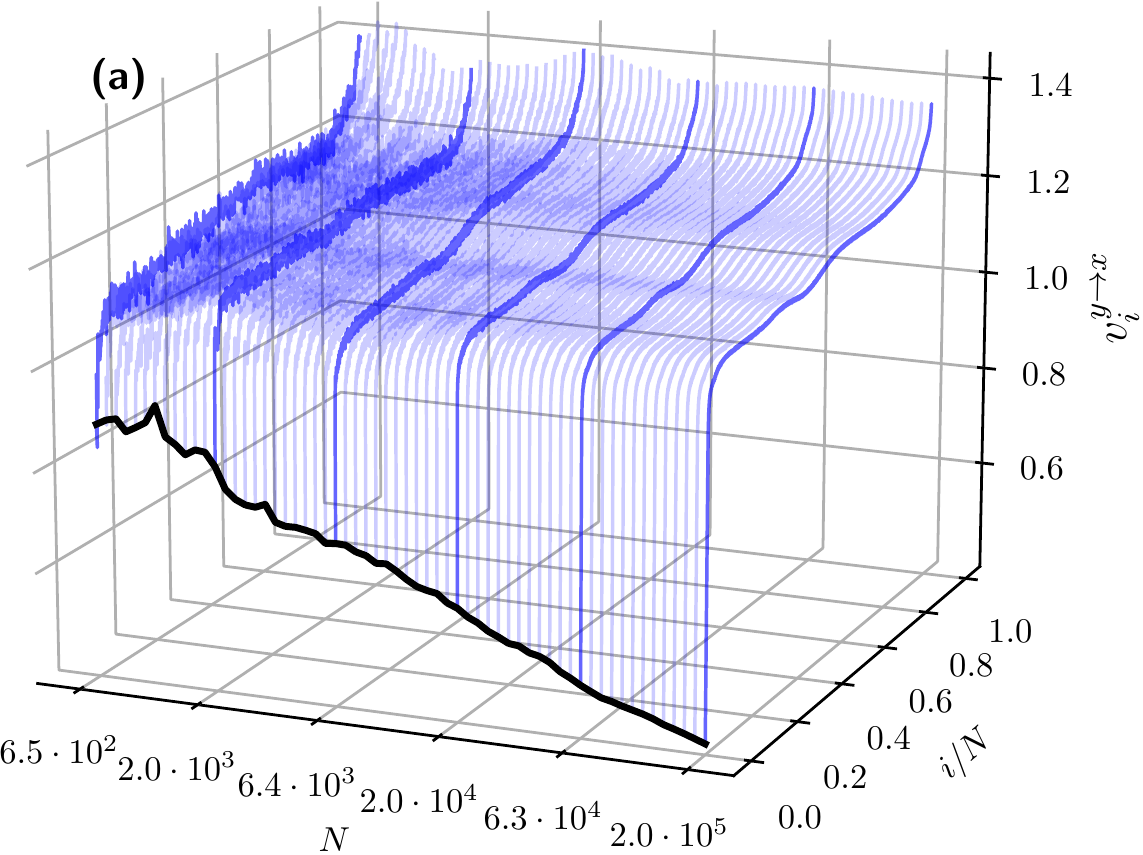}
\includegraphics[width = \linewidth]{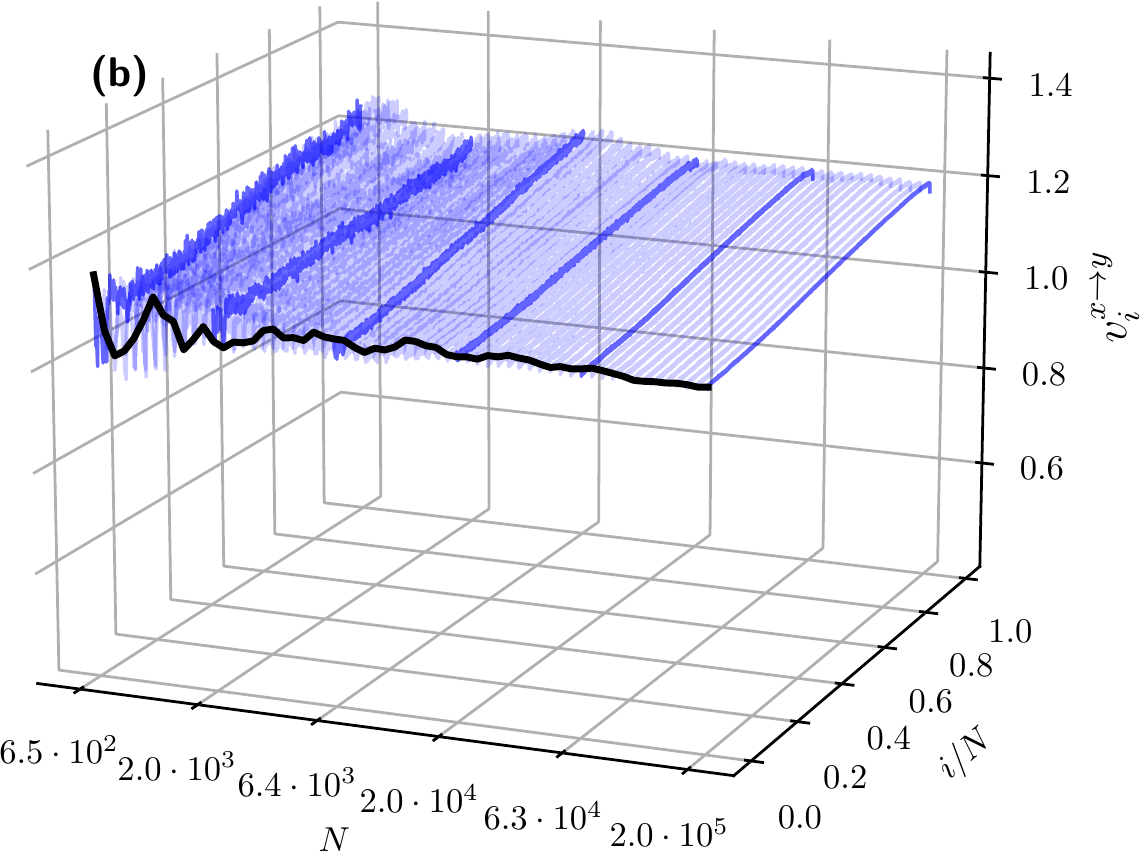}
\caption{
The dependence of the cross-distance vectors $\bm{v}^{y\rightarrow x}$ and $\bm{v}^{x\rightarrow y}$ on the length of the time series $N$. The blue lines represent the cross-distance vectors at a certain $N$ (these are the same lines as in, for example, \figurename\ref{figcouplingvectors}). The black lines represent the values of the first points of the cross-distance vectors. For a better visibility, six cross-distance vectors are highlighted and have $N$ that is written in the figure ticks. The scale on the $N$ axis is logarithmic. The $i$ axis of each plotted cross-distance vector was normalised to $i/N$ (a value between 0 and 1) for a simpler comparison.}
\label{Ndependence}
\end{figure}

An important property is the convergence of the cross-distance vectors with the length of the time series $N$.
In \figurename\ref{Ndependence}, the dependence of the cross-distance vectors on the length of the time series $N$ is shown at constant $L = 20$.


The beginning of the cross-distance vector $\bm{v}^{y\rightarrow x}$ is significantly smaller than the bulk, i.e., the vector has an initial tail.
By increasing $N$, the first point (represented by the black line) lowers even further from the bulk, which is desired.
This happens because any entry of the cross-distance vector $\bm{v}^{y\rightarrow x}$ cannot be smaller than the smallest entry of the distance matrix \eqref{eq:dist_y}, since the vector entry itself is an average of these distances.
When dealing with a finite number of segments (finite $N$), the smallest distance between a pair of segments will be a finite value.
With increased $N$, the smallest distance (most likely) decreases due to a bigger number of segments.
Thus, the lowest possible value of the cross-distance vector also decreases, and so do the initial values of $\bm{v}^{y\rightarrow x}$.
This means that coupling is easier to detect for longer time series.
In the limit, $N \rightarrow \infty$, the initial values of $\bm{v}^{y\rightarrow x}$ would reach zero.

In the opposite direction with no coupling, the effect is opposite.
The initial values of $\bm{v}^{x\rightarrow y}$ converge towards the bulk of the vector, i.e., there is no initial tail, which is also desired.
This means that at large enough $N$, no coupling is detected in the direction with no coupling.

Another behaviour we notice from \figurename\ref{Ndependence} is the change of the whole shape of the cross-distance vectors.
Their smoothness increases with increased $N$, and they seem to converge to a particular shape (which depends on the system).

To sum up, by increasing the length of the time series $N$, the reliability of this method increases.
For large enough $N$, the coupling will not be detected in the direction without coupling, and it will be detected in the direction of coupling.
Interestingly, the coupling direction is reliably inferred in short time series that contain only about 40 oscillations.

\subsection{Algorithm parameter dependence} \label{sectionalpardep}
\begin{figure}
\includegraphics[width = \linewidth]{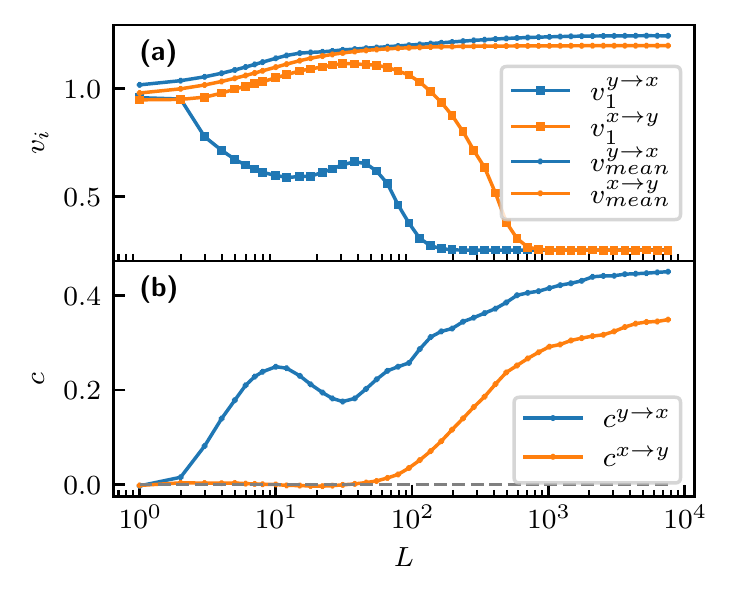}
\caption{The dependence of the first points and the averages of the cross-distance vectors (a) and the coupling indices (b) on the segment length $L$. Coupling indices were obtained with equation \eqref{Dequation} with $k_1 = 10$ and $k_2 = 100$.}
\label{Ldependence}
\end{figure}

We use the same test system for this analysis.
The time series length is $N = 2 \cdot 10^4$.
The only parameter of the cross-distance vectors algorithm is the segment length $L$.
\figurename \ref{Ldependence}~(a) shows the dependence of the first points and of the mean of the cross-distance vectors on the segment length $L$.

The value of the first point $v_1^{y\rightarrow x}$ generally decreases with increasing $L$ up to $L \approx 200$.
Since the mean values of both cross-distance vectors increase only by a little with increasing $L$, this indicates that the significant change is in the initial tail.
In the other direction, $v_1^{x\rightarrow y}$ stays close to the mean $v_{\text{mean}}^{x\rightarrow y}$ until it starts to decrease at around $L \approx 50$.
This is reflected in the coupling indices $c^{y\rightarrow x}$ and $c^{x\rightarrow y}$, the dependence of which is shown in \figurename \ref{Ldependence}~(b).
It tells us two important properties.


First, we notice that $c^{y\rightarrow x}$ does not detect coupling for $L = 1$.
This is because, at $L = 1$, the subsystems are not well reconstructed with the segments.
The Takens' embedding theorem \cite{Takens1981} gives a minimum dimension of delay embedding vectors needed for reconstructing a system's attractor.
It is $2n+1$, where $n$ is the system dimension (though often, less than that is needed).
For Duffing subsystems in our example, $L=1$ does not reconstruct the attractor, which results in falsely not detecting coupling.

Second, we can see that for $L > 1$, $c^{y\rightarrow x}$ is significantly larger than $c^{x\rightarrow y}$ and generally increases with increased $L$, except for a local maximum at $L=10$.
The other index $c^{x\rightarrow y}$ stays close to zero until $L\approx 50$, where it starts to increase visibly.
This means that a too large $L$ results in falsely detecting coupling.
The inherent similarity of neighbouring segments is the reason for false coupling detection in the direction $x\rightarrow y$ at $L > 50$.

The similarity between segments will generally decrease if we increase segment length $L$.
This becomes obvious when we notice that with increased $L$, more dimensions of segments must match to maintain high similarity.
However, the neighbouring segments $\bm{o}^x_i$ and $\bm{o}^x_{i+1}$ are autocorrelated and will, therefore, always be very similar for any $L$.
This can be seen from the definition of the distance measure \eqref{distanceequation}.

Therefore, when $L$ becomes large enough, the closest segment to $\bm{o}_i^x$ will most likely be $\bm{o}_{i+1}^x$ (or $\bm{o}_{i-1}^x$).
The same will hold for segments of the other subsystem, i.e., the closest to $\bm{o}_i^y$ will most likely be $\bm{o}_{i+1}^y$.
This makes it seem as if both statements \eqref{statement1} and \eqref{statement2} are technically correct (but only for time autocorrelated segments), regardless of whether there is coupling.
This is the reason for the increase of $c^{x\rightarrow y}$ at $L \approx 50$ in our example.
It shows that $L$ cannot be too large as it can result in falsely detecting coupling.
A more detailed analysis of this effect is done in Appendix~\ref{appendixnl}.
Guidelines for tuning this parameter are given in Section~\ref{sectioncompdetails}.


The detected coupling direction turns out to be robust with respect to the choice of $k_1$ and $k_2$ when calculating the coupling indices.
The general rule is that $k_1$ should contain the initial tail, which contains the information about coupling.
The choice of $k_2$ is significantly less important.
In this article, we chose $10k_1 < k_2 < 100k_1$, though smaller and larger values give similar results.
The point is that $k_2$ must be much smaller than $N$.
An analysis of the influence of $k_1$ and $k_2$ values on the indices $c$ is done in Appendix~\ref{appendixk}.

\subsection{Noise dependence}\label{secnoise}
Let us consider the noise robustness of this method.
We use time series of the same test system.
The time series length is  $N = 5\cdot 10^4$.
We add Gaussian noise to each point of the time series $x(t_i) \rightarrow x(t_i) + \xi_i, \xi_i \sim \mathcal{N}(0, \sigma^2)$, where $\sigma$ is the standard deviation of the noise.

\begin{figure}
\includegraphics[width = \linewidth]{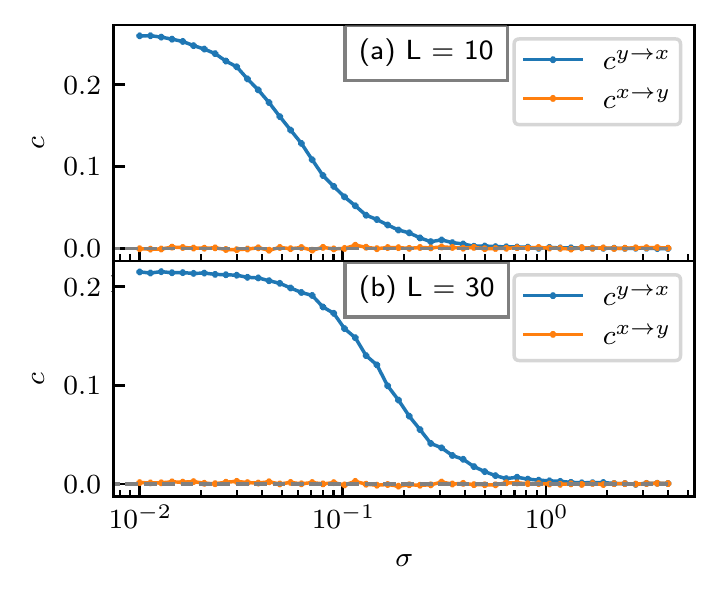}
\caption{The noise dependence of the coupling indices $c^{y\rightarrow x}$ and $c^{x\rightarrow y}$ with segment lengths $L = 10$ (a) and $L = 30$ (b).
Coupling index parameters are $k_1 = 10$ and $k_2 = 100$.}
\label{noisedependence}
\end{figure}

In \figurename\ref{noisedependence}, the dependence of the coupling indices $c^{y\rightarrow x}$ and $c^{x\rightarrow y}$ on $\sigma$ is shown.
As one might expect, the reliability of the detected coupling direction decreases with increased $\sigma$.
For $L = 10$, the index $c^{y\rightarrow x}$ decreases to zero at around $\sigma \approx 0.4$, and for $L = 30$ at around $\sigma \approx 0.7$, which is more than a third of the subsystems' amplitudes.
The other index $c^{x\rightarrow y}$ is close to zero for all values of $\sigma$.

Noise robustness can be further increased by increasing $L$.
However, one must be careful not to increase it to the point of false detection, explained in Section \ref{sectionalpardep}.
This method is, therefore, quite robust to noise.

\subsection{Robustness to artefacts and missing data}

Measurements can contain artefacts, such as spikes, or have missing values for a time period.
Coupling detection in such data can be problematic.
Coupling indices are resilient to such imperfections in time series.

The solution is to delete the distance matrix elements whose values were obtained from segments with artefacts or missing values.
This is done by deleting the corresponding rows and columns, decreasing the matrix size.
They must be deleted from both distance matrices, even if the artefact is only present in one time series.
This allows us to ignore any unwanted sections of either time series, making this approach very flexible.
The only requirement is manually choosing the time series points to ignore.

To give an example, consider the time series of the same test system.
The time series length is  $N = 5\cdot 10^4$.
We change the values of $x$ between points $5000$ and $9000$ and the values of $y$ between $30000$ and $36000$ to random values to simulate artefacts.
Therefore, when computing the distance matrices, we ignore $4000+6000=10000$ rows and columns (which will result in cross-distance vectors with $10000$ fewer points).
Let us compare cross-distance vectors, $c$ indices, and $M$ indices obtained from clean time series and time series containing artefacts.

\begin{figure}
\includegraphics[width = \linewidth]{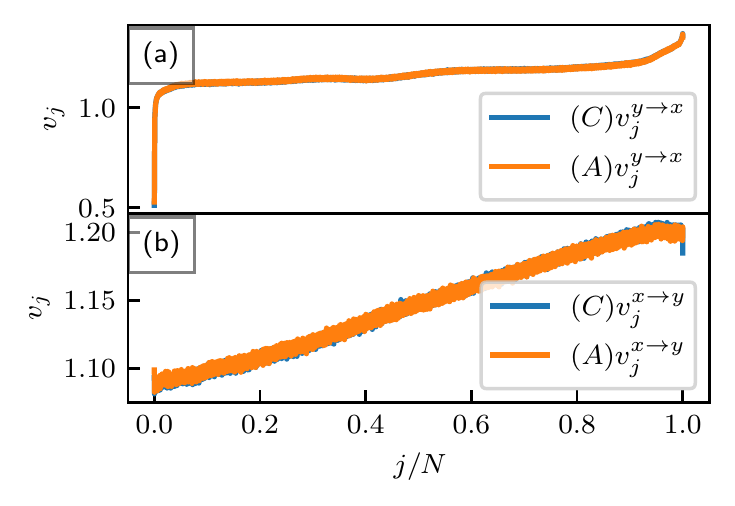}
\caption{Cross-distance vectors obtained from clean data (C) and from data with artefacts (A).
$\bm{v}^{y\rightarrow x}$ is shown in (a) and $\bm{v}^{x\rightarrow y}$ is shown in (b).
The $j$ axis of each plotted cross-distance vector was normalised to $j/N$ (a value between 0 and 1) for a simpler comparison.
Note that the vectors obtained from clean and from data with artefacts are nearly indistinguishable.
The index values obtained from clean data are 
$c^{y\rightarrow x} = 0.268$, $c^{x \rightarrow y} = 0.000$, $M(Y|X) = 0.407$, $M(X|Y) = 0.052$.
The index values obtained from data with artefacts are 
$c^{y\rightarrow x} = 0.263$, $c^{x \rightarrow y} = -0.001$, $M(Y|X) = 0.385$, $M(X|Y) = 0.049$.}
\label{artefacts}
\end{figure}

The results are shown in \figurename~\ref{artefacts}.
The cross-distance vectors obtained from clean data and data containing artefacts are nearly indistinguishable.
The $c$ and the $M$ indices are also nearly the same.
The difference in $c$ is seen in the third decimal.
The difference in $M$ is larger but still small.
This shows that state space approaches are resilient to artefacts and missing values.

\subsection{Practical computational details}
\label{sectioncompdetails}

In this section, we will briefly discuss the practical aspects of the algorithm and provide a suggestion that can prove useful.


\paragraph*{Tuning the L parameter}
For this purpose, we can use the large similarity of neighbouring segments (explained in Section \ref{sectionalpardep}).
For a chosen $L$, we can check whether the closest segment is often one of the neighbours, i.e., whether the distance matrices \eqref{eq:dist_x} and \eqref{eq:dist_y} have very small subdiagonals.
If at least one of them does, we must lower $L$ until subdiagonals have similar values to the rest of the matrices since small subdiagonals lead to false detection.
This gives the maximal possible value $L_{max}$.
In practice, we recommend a value close to the maximal $L_{max}$, for example, $L_{max}/2$, since larger $L$ generally gives more accurate and noise-robust results.


\paragraph*{Neighbouring segments}

In some cases, especially in time series with a large sampling rate (small time step~$\Delta t$), lowering $L$ cannot adequately raise the values of the subdiagonals.
If the time step is very small, the neighbouring segments will be autocorrelated and will therefore always be very close such that $d(\bm{o}_i^x, \bm{o}_{i+1}^x) \rightarrow 0$.
This can lead to false coupling detection, as explained in \ref{sectionalpardep}.
We propose two solutions.

The first solution is to downsample the signal.
The neighbouring segments become less similar by increasing the time step $\Delta t$.
We must downsample the time series to the point where neighbouring segments are no longer the most similar, i.e., when the subdiagonal values of the distance matrices are no longer small compared to the rest of the values.

The second solution is manually decreasing the dimensions of the distance matrices.
This is needed in a case where downsampling is not an option.
One such example is stiff subsystems, i.e., subsystems which contain small and large frequency components.
In such cases, downsampling can erase the high-frequency component (and thus erase possibly crucial information) but still not raise the values of the subdiagonals.
For this purpose, we suggest an alternative approach.
We can only calculate every $M$-th value of the full distance matrices, i.e., we construct them from every $M$-th segment.
Choosing $M$ large enough ignores the (small) subdiagonals that appear in full matrices.

Both approaches should be done with a fixed $L$, for example, $L = 5$.
Once the subdiagonals are adequately raised, the chosen approach can be repeated for a larger $L$.
If successful, the larger $L$ should be taken for better accuracy and noise robustness.

Both approaches ignore small subdiagonals that would appear in full matrices, as well as other values that can be redundant.
For example, a full matrix contains elements $d(\bm{o}_i^x, \bm{o}_{j}^x)$ and $d(\bm{o}_i^x, \bm{o}_{j+1}^x)$, which have similar values if $d(\bm{o}_j^x, \bm{o}_{j+1}^x) \approx 0$.

The downsampling approach is computationally more efficient, as it effectively decreases segment length $L$.
Therefore, if possible, it should be chosen over manually decreasing dimensions.

\paragraph*{Time and memory limitations}
In a time series of length $N$, the computational complexity of the algorithm is {$\mathcal{O}(N^2\log(N))$ due to the sorting of all rows of the distance matrices.
As the previous paragraph explains, downsampling the original time series is recommended since it significantly decreases the execution time.

Furthermore, the matrices $\bm{D}^x$ and $\bm{D}^{y\rightarrow x}$ can, in practice, be too large to store in computer memory.
To avoid this, the rows of the matrices \eqref{eq:dist_x} and \eqref{eq:dist_y} can be calculated individually to obtain a single row of $\bm{D}^{y\rightarrow x}$.
One summation in \eqref{couplingvector} is done for each row, and the cross-distance vector is obtained by only storing a few sets of data of size $N-L+1$ at once.

\paragraph*{GPU implementation} 


Since the algorithm is based on matrix operations, GPU devices can significantly decrease the execution time.
For this purpose, we are providing a GPU-based implementation written in Python using the JAX library~\cite{jax}.
A time-efficient and memory-efficient implementation is available at \cite{repo}.

\section{Application to physiological signals}
\label{sectiobio}

In this section, we will apply the cross-distance vectors algorithm to time series of physiological measurements.
The goal is to show the applicability of cross-distance vectors to real-life systems.
We will analyse coupling in the human cardiorespiratory system.
The cardiac subsystem is characterised by ECG (electrocardiogram), and the respiratory subsystem by respiration curves obtained with a respiratory belt sensor.
These subsystems are inherently bidirectionally coupled, i.e., both subsystems depend on each other.
A survey of the mechanisms responsible for this dependence is given in \cite{Dick2014}.
Specifically, we will investigate coupling in the direction from the respiratory to the cardiac subsystem, which we label with~$R\rightarrow C$.

The signals used in the analysis are shown in \figurename\ref{biosignals}.
Thirty minutes of ECG and respiration curves were measured on two subjects.
In such a way, we obtain two pairs of signals, marked $R_1$ and $C_1$ for the first subject and $R_2$ and $C_2$ for the second subject. 
\begin{figure}
\includegraphics[width = \linewidth]{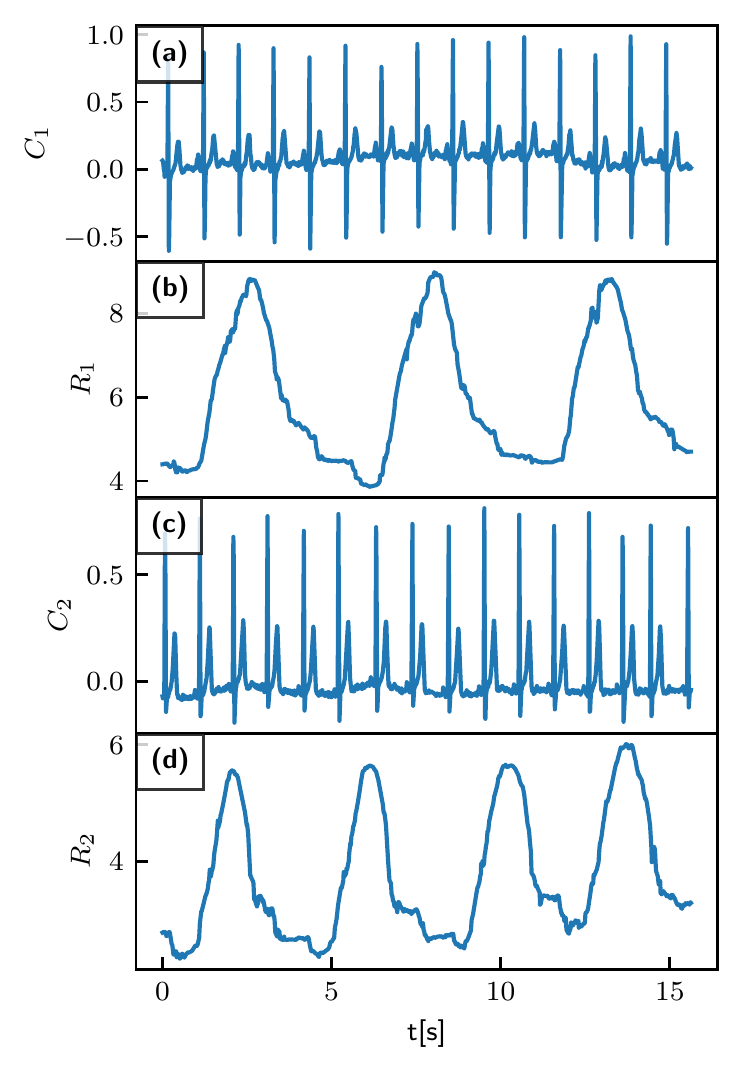}
\caption{A part of the ECG and respiration signals used for analysis.
Label $C$ means ECG, and label $R$ means respiration. $R_1$ and $C_1$ were measured simultaneously on one subject, and $R_2$ and $C_2$ simultaneously on another subject. Time series parameters are $N = 172800$ and $\Delta t = \frac{1}{96}s$, corresponding to a measurement time of $30$ minutes.}
\label{biosignals}
\end{figure}

The most dominant frequency component in ECG signals is around 1Hz.
The R peak is, however, very short, which means this subsystem is stiff.
Therefore we do not downsample the time series but rather manually decrease the distance matrix sizes by a factor $M$, as explained in Section~\ref{sectioncompdetails}.
In this case, we chose $M = 20$, which means we are comparing segments that are apart by a multiple of $M \Delta t = 20 / 96s \approx 0.2s$.

\figurename\ref{biocoupling} (a) and (b) represent cross-distance vectors for $R_1 \rightarrow C_1$ and $R_2 \rightarrow C_2$.
In these cases, ECG and respiratory signals belong to the same person and were measured simultaneously.
Therefore, we expect to detect coupling.
Indeed, the cross-distance vectors have an initial tail, suggesting coupling is present in the underlying subsystems.

\figurename\ref{biocoupling} (c) and (d) represent cross-distance vectors for $R_2 \rightarrow C_1$ and $R_1 \rightarrow C_2$.
Since the signals belong to different subjects that are inherently independent, we do not expect to detect coupling.
Indeed, the cross-distance vectors do not have an initial tail, suggesting no coupling in the underlying subsystems.

Let us compare the $c$ indices, obtained with parameters $k_1 = 10, k_2 = 1000$.
The indices  $c^{R_1 \rightarrow C_1} = 0.105$ and $c^{R_2 \rightarrow C_2} = 0.176$ are nearly two orders of magnitude larger than $c^{R_2 \rightarrow C_1} = -0.0004$ and $c^{R_1 \rightarrow C_2} = -0.006$.
This means that coupling indices have significantly larger values when the coupling is present in the underlying subsystems than when the subsystems are independent.
Coupling indices $c^{R_2 \rightarrow C_1}$ and $c^{R_1 \rightarrow C_2}$ even have a small negative value, which is a strong indication of the absence of coupling.

For comparison, the $M$ indices were also computed with parameter $k = 10$.
They were computed from the same distance matrices as the $c$ indices, i.e., matrices obtained by comparing every 20th segment with $L = 20$.
The obtained values are $M(R_1|C_1) = 0.150$, $M(R_2|C_2) = 0.254$, $M(R_2|C_1) = -0.003$, and $M(R_1|C_2) = -0.009$.
The $M$ indices also correctly detect the existence of coupling in the first two cases and have a small negative value in the second two cases.

When considering the coupling indices $c$ in an application, it should be noted that they require the time series to belong to a dynamical system that can be reconstructed with time-delay embedding.
Otherwise, approaches such as Granger causality or information transfer are more suitable.
However, contrary to the indices $c$, these approaches may not be reliable in short time series.
To sum up, the new coupling indices $c$ excel in causality detection from bivariate time series generated by a dynamical systems, especially in short signals.

\begin{figure}
\includegraphics[width = \linewidth]{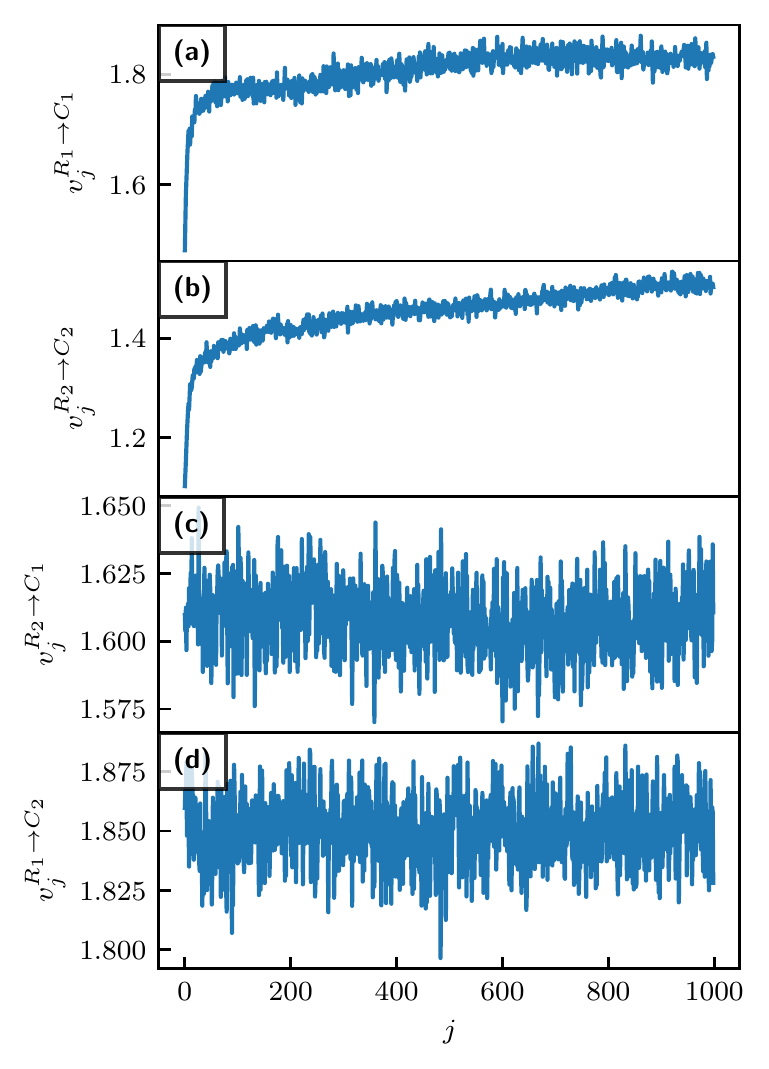}
\caption{The first 1000 points of the cross-distance vectors in the direction from the respiratory to the cardiac subsystem.
In (a) and (b), the analysed signals belong to the same subject, while in (c) and (d), they are independent.
The segment length is $L = 20$.
Dimensions of distance matrices were decreased by a factor $20$.}
\label{biocoupling}
\end{figure}

\section{Conclusion}

In this article, we have defined \textit{cross-distance vectors} as a means of inferring the direction of coupling from bivariate time series and provided an algorithm for calculating them.
Cross-distance vectors provide information about coupling by calculating two coupling indices, which quantify coupling strength in both directions.

The new coupling indices can infer the coupling direction in various coupled dynamical systems.
Comparing the new coupling indices to the conventional indices based on state space distance shows more accurate results with the new approach.
Analysis of numerical stability has shown that the reliability of the new indices increases with the length of the time series but is also reliable in short time series, containing only about 40 oscillations.
The performance of the coupling indices was also evaluated in detecting cardiorespiratory interaction in measured data.

The algorithm has a simple implementation that requires the choice of only a maximum of three parameters.
The selection of the optimal parameters' values can be made systematically, which results in robust performance and circumvents the challenge of determining optimal embedding parameters.
A numerically efficient implementation is available at \cite{repo}.

A logical continuation is an extension of the new method for inferring connections in a network of multiple subsystems.
Should the multivariate extension be as effective as the bivariate variant, it could become an essential tool in analysing complex multivariate problems such as brain connectivity.

\appendix

\section{Mapping of close states}
\label{appendix}

In what follows, we will provide an argument to illustrate why close states in the driven subsystem are mapped to close states in the driving subsystem for various subsystems with different properties.

One might assume that the statement \eqref{statement1} is false and the statement \eqref{statement2} is true.
It may seem that the times when the driving subsystem is self-similar will appear as times of large self-similarity in the driven subsystem via coupling.
Hence \eqref{statement2} is true.
Since this logic does not apply in the other direction, one might expect that the statement \eqref{statement1} is false.
This is, however, not the case.
In a special case under two assumptions, we provide an analytic argument.
For a more general case, we will provide a heuristic argument.

\paragraph*{Special case}
Consider that two segments of the driven subsystem $\bm{o}_i^x$ and $\bm{o}_j^x$ are identical (this is only possible for subsystems in a regular dynamical regime)
\begin{align}
& \bm{x}(t_{i+k}) = \bm{x}(t_{j+k}), \label{xequality} \\
& k = 0,1, \dots, L-1 \nonumber.
\end{align}
If the subsystem coordinates match over a time period, their time evolution must also match.
The first assumption is that $\bm{f}$ is time-independent, which gives us the following $L$ equations
\begin{equation}
\begin{split}
& \bm{f}(\bm{x}(t_{i+k})) + \bm{g}(\bm{x}(t_{i+k}),\bm{y}(t_{i+k})) = \\
& \bm{f}(\bm{x}(t_{j+k})) + \bm{g}(\bm{x}(t_{j+k}),\bm{y}(t_{j+k})), \\
& k = 0,1, \dots, L-1.
\label{equation2}
\end{split}
\end{equation}
Additionally, from \eqref{xequality} follows
\begin{align}
& \bm{f}(\bm{x}(t_{i+k})) = \bm{f}(\bm{x}(t_{j+k})) \label{fequality}\\
& k = 0,1, \dots, L-1 \nonumber.
\end{align}
Combining \eqref{equation2} and \eqref{fequality} gives us
\begin{align}
& \bm{g}(\bm{x}(t_{i+k}),\bm{y}(t_{i+k})) = \bm{g}(\bm{x}(t_{j+k}),\bm{y}(t_{j+k})), \label{gequality} \\
& k = 0,1, \dots, L-1 \nonumber.
\end{align}
Since \eqref{xequality} holds, we can consider the first argument of $\bm{g}$ as constant and define $k$ functions
\begin{align}
& \tilde{\bm{g}}(\bm{y}(t_{i+k})) = \bm{g}(\bm{x}(t_{i+k}),\bm{y}(t_{i+k})), \\
& k = 0,1, \dots, L-1 \nonumber.
\end{align}
The second assumption is that functions $\tilde{\bm{g}}(\bm{y}(t_{i+k}))$ are injective.
If this is the case, from \eqref{gequality} follows
\begin{align}
& \bm{y}(t_{i+k}) = \bm{y}(t_{j+k}), \label{yequality} \\
& k = 0,1, \dots, L-1 \nonumber.
\end{align}
This means that statement \eqref{statement1} is \textbf{true}.
This argument does not work in the other direction with swapped $\bm{x}$ and $\bm{y}$, since $\bm{y}$ does not depend on $\bm{x}$, i.e., there is no coupling function in the time evolution of $\bm{y}$.
Therefore, the statement \eqref{statement2} is \textbf{false}.

\paragraph*{General case}
We argue that if the subsystems are sufficiently nice, this also holds without the two assumptions if segment length $L$ is large enough.
To obtain \eqref{gequality} from \eqref{equation2} with time dependent $\bm{f}(\bm{x},t)$, we argue that it would seem unlikely for the sum of $\bm{f}$ and $\bm{g}$ to match over a time period, if they do not match individually.
The analysis in Section \ref{ssduffing} shows that this indeed holds for coupled periodically forced Duffing subsystems.
To obtain \eqref{yequality} from \eqref{gequality} if $\tilde{\bm{g}}(\bm{y}(t_{i+k}))$ are not injective, we would similarly argue that if $\bm{g}$ matches over a long enough time period, so must its arguments.

\bigskip

We can use the same arguments for chaotic subsystems by demanding the distance of segments to be less than $\delta$ instead of them being identical.
Under the admissibility conditions of the Poincaré recurrence theorem, the choice of $\delta$ can be arbitrarily small.
The arguments hold for chaotic subsystems by swapping all the equalities in equations \eqref{xequality}--\eqref{yequality} with arbitrarily small proximity.

\section{Detailed comparison of $c$ and $M$}
\label{appendixB}

Here the goal is to compare the numerical stability of $c$ and $M$ indices.
For that purpose, the analysis from Section~\ref{secnumstab} is repeated for $M$, and both results are presented jointly.
All the analysis is done on the same test system of coupled Duffing oscillators as in Section~\ref{secnumstab}.
Also, all the parameters are the same as in Section~\ref{secnumstab}.
Coupling is unidirectional with direction $y\rightarrow x$.

\subsection{Dependence on time series length}

The analysis in Section~\ref{secNdependence} was done on the cross-distance vectors.
For results to be comparable to $M$, we compute the $c$ indices.
As explained in Section~\ref{seccvsm}, the indices are most fairly compared when their parameters are $k = k_1$.
In our case, $k=k_1 = 10$ and $k_2 = 100$ are taken, the same as in Section~\ref{ssduffing}.
While these are not optimal parameters for every $N$, we keep them fixed for simplicity.
The segment length is $L=20$, as in Section~\ref{secNdependence}.
The comparison is done in \figurename~\ref{MNdependence}.
\begin{figure}
\includegraphics[width = \linewidth]{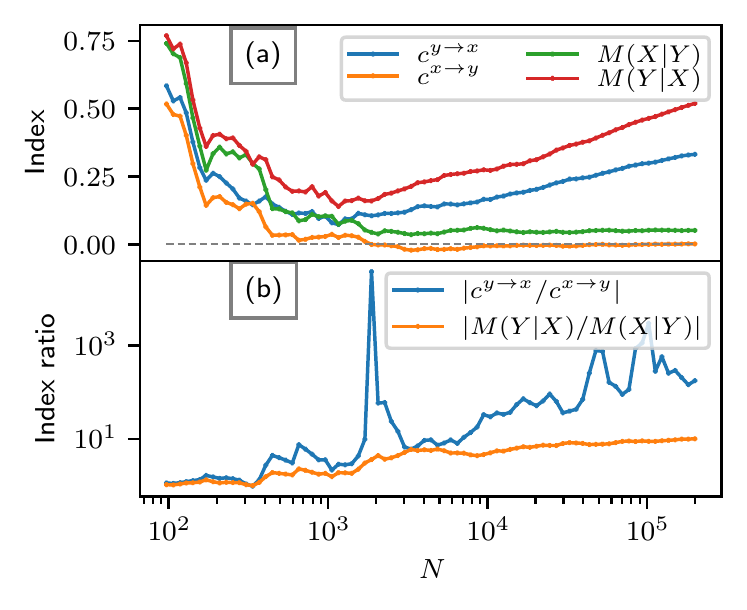}
\caption{The dependence of the indices $c^{y\rightarrow x}$, $c^{x\rightarrow y}$, $M(Y|X)$, $M(X|Y)$ (a) and the index ratios $c^{y\rightarrow x}/c^{x\rightarrow y}$, $M(Y|X)/M(X|Y)$ (b) on the length of the time series $N$.
The $M$ index parameter is $k=10$ and the $c$ index parameters are $k_1=10, k_2=100$.}
\label{MNdependence}
\end{figure}

The general behaviour is similar for both indices, as seen in \figurename~\ref{MNdependence}(a).
At small $N$, all indices are large.
The main difference between $M$ and $c$ is seen at large $N$ where $c^{x\rightarrow y}$ converges to zero, which is desirable, while $M(X|Y)$ seems to converge to a finite positive value.
As explained in Section~\ref{couplingindexsection}, this is because $c$ ignores possible trends in the cross-distance vectors, while $M$ does not.

The index ratios are shown in \figurename~\ref{MNdependence}(b).
Ideally, the ratios are infinite since the coupling is unidirectional.
For nearly all values of $N$ the ratio $c^{y\rightarrow x}/c^{x\rightarrow y}$ is larger than the ratio $M(Y|X)/M(X|Y)$, especially for large $N$.
For small $N$, the difference is smaller, but the $c$ ratio is generally still larger.
This shows that $c$ better determines the coupling direction for both short and long time series.

\subsection{Dependence on $L$}

Let us compare the $L$ dependence from \figurename~\ref{Ldependence} for indices $c$ and $M$.
The chosen time series length is $N = 2 \cdot 10^4$, same as in Section~\ref{sectionalpardep}.

The dependence of the indices $c$ and $M$ on the segment length $L$ is shown in \figurename~\ref{MLdependence}(a).
The dependence is very similar for both index variants.
The main difference is that both $M$ indices are slightly larger (since they do not ignore trends seen in cross-distance vectors).
At $L=1$, the $M$ indices both have a very similar positive value, while the $c$ indices are both much closer to zero.
Interestingly, the bias of $M(X|Y)$ due to trends is nearly constant for segment length up to $L\approx 50$, where $L$ becomes too large (as explained in Section~\ref{sectionalpardep}).

The index ratios $c^{y\rightarrow x}/c^{x\rightarrow y}$ and $M(Y|X)/M(X|Y)$ are shown in \figurename~\ref{MLdependence}(b).
The ratio of the $c$ indices is significantly larger than that of the $M$ indices for all values of $L$, except for $L=1$.
This means that the $c$ indices better determine the coupling direction regardless of the choice of segment length $L$.

\begin{figure}
\includegraphics[width = \linewidth]{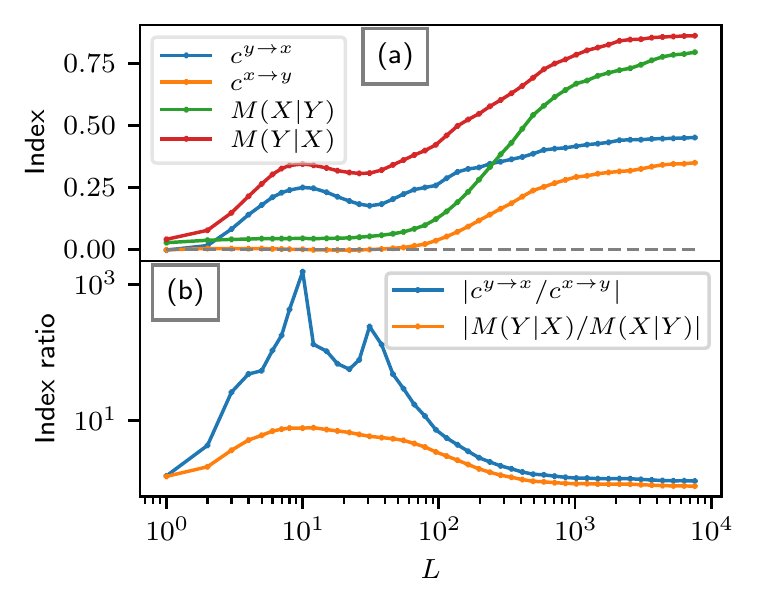}
\caption{The dependence of the $c$ and $M$ indices (a) and the index ratios $c^{y\rightarrow x}/c^{x\rightarrow y}$ and $M(Y|X)/M(X|Y)$ (b) on the segment length $L$.
The $M$ index parameter is $k=10$ and the $c$ index parameters are $k_1=10, k_2=100$.}
\label{MLdependence}
\end{figure}

\subsection{Dependence on noise}

Let us compare the noise robustness of the $c$ and the $M$ indices similarly as in Section~\ref{secnoise}.
The chosen time series length is again $N = 5 \cdot 10^4$ and the chosen segment length is $L = 10$, the same as in \figurename~\ref{noisedependence}(a).

The dependence of the $c$ and $M$ indices on the standard deviation of noise $\sigma$ is shown in \figurename~\ref{Mnoisedependence}(a).
It turns out that the dependence is very similar for both indices.
They are both robust to noise.
As seen in \figurename~\ref{Mnoisedependence}(b), the index ratio is larger for $c$ than for $M$ up to around $\sigma \approx 0.5$, at which point both ratios become close to $1$.


\begin{figure}
\includegraphics[width = \linewidth]{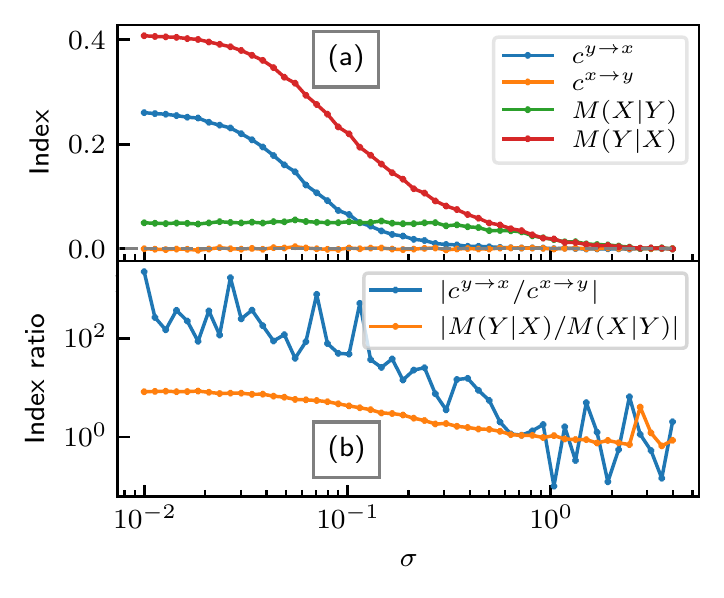}
\caption{The dependence of the $c$ and $M$ indices (a) and the index ratios $c^{y\rightarrow x}/c^{x\rightarrow y}$ and $M(Y|X)/M(X|Y)$ (b) on the standard deviation of noise $\sigma$.
The $M$ index parameter is $k=10$ and the $c$ index parameters are $k_1=10, k_2=100$.
The segment length is $L = 10$.}
\label{Mnoisedependence}
\end{figure}

\section{The $N$ and $L$ dependence of cross-distance vectors}
\label{appendixnl}


In order to get a complete picture of the behaviour of the cross-distance vectors, we compute them for numerous values of pairs $(N,L)$.
We again use the test system from Section~\ref{secnumstab}.
We are interested in the cross-distance vectors' first (nonzero) point.
The results are shown in \figurename~\ref{NLdependence}.
\figurename~\ref{NLdependence} can be understood as plotting the black lines in \figurename~\ref{Ndependence}, calculated for different $L$.
Equivalently, \figurename~\ref{NLdependence}(a) can be seen as plotting $v_1^{y\rightarrow x}$ from \figurename~\ref{Ldependence}(a) and \figurename~\ref{NLdependence}(b) as $v_1^{x\rightarrow y}$ from \figurename~\ref{Ldependence}(a), both calculated for different $N$.

\figurename~\ref{NLdependence}(a) shows $v_1^{y\rightarrow x}$, which corresponds to the direction of coupling.
One can roughly identify three areas in the $(N,L)$ grid that are highlighted using two red lines.
Mind that the areas are not strictly defined but assist in explaining the figure.
Area 1 has very small values.
There, $L$ is too large at a given $N$, which results in detecting coupling regardless of the underlying dynamics (area of falsely detecting coupling).
Area 3 has large values.
There $L = 1$, which is too small (at any $N$) to reconstruct the underlying state space, which results in not detecting coupling regardless of the underlying dynamics (area of falsely not detecting coupling).
Area 2 has moderate values.
There, $L$ is large enough to reconstruct the underlying state space but small enough to avoid false detection.

The white line highlights the approximate area where the optimal $L$ at a given $N$ is.
We understand the optimal $L$ as the value in area 2 where $v_1^{y\rightarrow x}$ is the smallest at a given $N$.
The optimal $L$ increases with increased $N$.

\figurename~\ref{NLdependence}(b) shows $v_1^{x\rightarrow y}$, which corresponds to the direction without coupling.
In this case, the three areas and the optimal $L$ are not as obvious as in the other direction.
We only clearly see where $L$ becomes too large, which leads to false coupling detection.

It should be noted that in both figures, there is a white area in the upper left corner.
This areas represents the impossible pairs $(N,L)$, i.e., $L\ge N$.

\begin{figure}
\includegraphics[width = \linewidth]{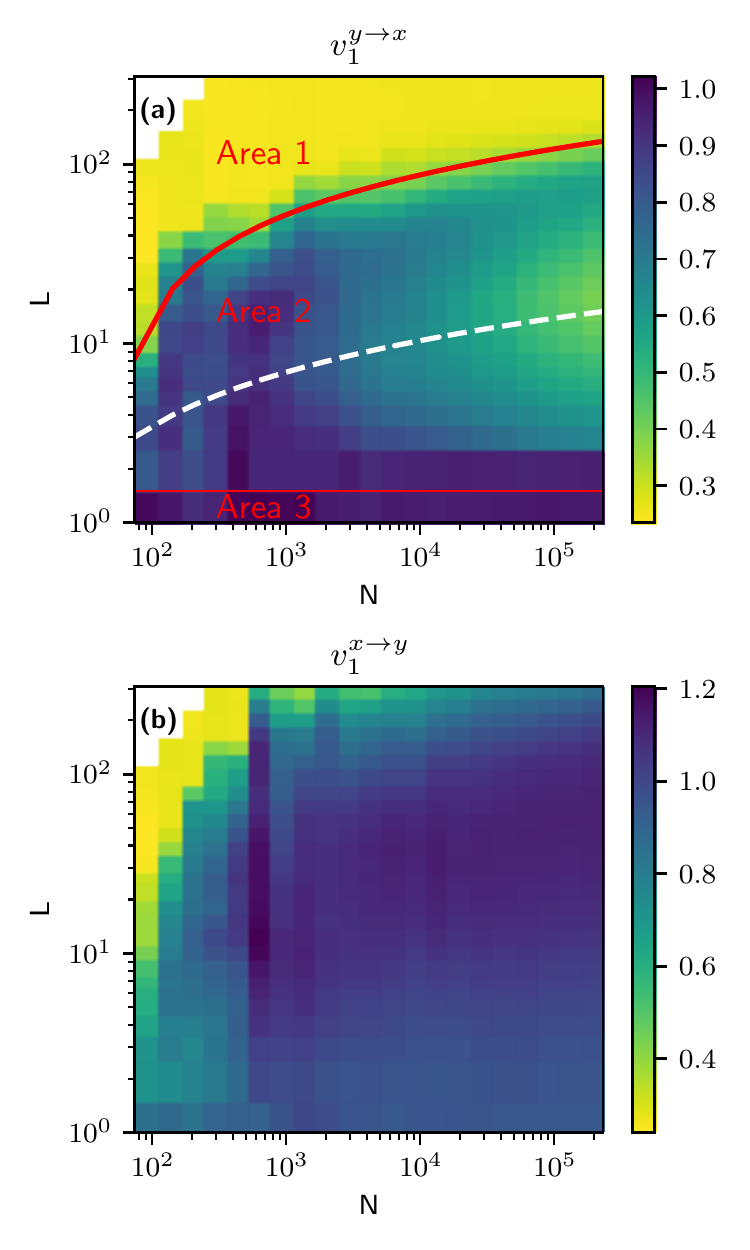}
\caption{The first point of the cross-distance vectors $v_1^{y\rightarrow x}$ (a) and $v_1^{x\rightarrow y}$ (b) for different values of the time series length $N$ and the segment length $L$.
In (a), the two red lines separate the grid into three areas, and the white line approximately represents optimal $L$ at a given $N$.
The white area in the upper left corner represents the impossible pairs $(N,L)$ where $L\ge N$.}
\label{NLdependence}
\end{figure}

\section{The choice of parameters $k_1$ and $k_2$}
\label{appendixk}


In order to analyse the influence of the parameters $k_1$ and $k_2$ on the coupling indices $c$, we consider the same test system used in Section~\ref{secnumstab}.
Specifically, we will plot the $c$ indices from \figurename~\ref{duffing1D}(a) (blue and orange lines), obtained with different $k_1$ and $k_2$.

The influence of the change of $k_1$ is shown in \figurename~\ref{k1dependence}.
In this case, we fixed the parameter $k_2 = 150$, which is close to $k_2$ most often used in the article.
We can make some important observations.
First, we notice that for small values of $k_1$, the values of $c^{y\rightarrow x}$ increase quickly with increased $\epsilon$, which allows for more reliable detection at small coupling.
Furthermore, the values of $c^{y\rightarrow x}$ reach higher values, while the values of $c^{x\rightarrow y}$ remain unchanged, which is also desirable.
However, there is a negative side to using such small values of $k_1$.
The variance of both indices is not negligible, as seen by a larger spread of $c^{x\rightarrow y}$ around zero and a negative value of $c^{y\rightarrow x}$ at $\epsilon=0$.
On the other hand, increasing $k_1$ decreases the variance of both indices.
However, it also reduces the rate of increase of $c^{y\rightarrow x}$ and introduces a bias at moderate values of $\epsilon$ (around $0.5$).
Therefore, the choice of $k_1$ offers a tradeoff between bias and variance of the indices~$c$.

The influence of the change of $k_2$ is shown in \figurename~\ref{k2dependence}.
Similar to the influence of $k_1$, the variance of both indices is larger at smaller $k_2$.
Furthermore, larger $k_2$ increases $c^{x\rightarrow y}$ at moderate values of $\epsilon$, increasing the false positive error.
Most importantly, this false positive error increases drastically when $k_2$ is close to the time series length $N$ ($k_2=49000$ in our case).
One can notice a strong similarity between the case of $k_2 = 49000$ and the $M$ indices in \figurename~\ref{duffing1D}(a) (green and red lines).
Based on this analysis, we draw the conclusion that the safe choice for $k_2$ is approximately $10k_1 < k_2 < 100k_1$.

\begin{figure}
\includegraphics[width = \linewidth]{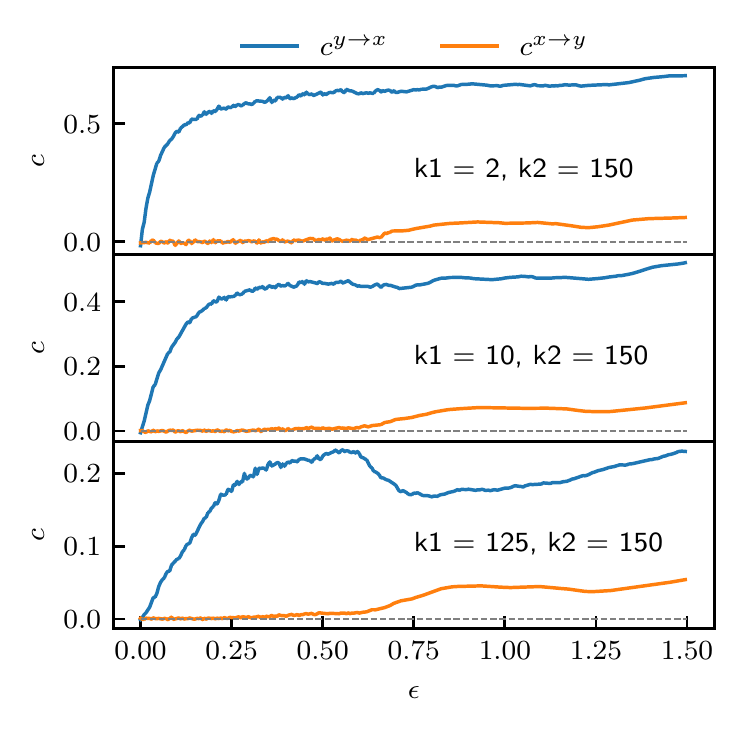}
\caption{The dependence of the coupling indices $c$ on the coupling parameter $\epsilon$ for different values of the parameter $k_1$ and fixed $k_2=150$.}
\label{k1dependence}
\end{figure}

\begin{figure}
\includegraphics[width = \linewidth]{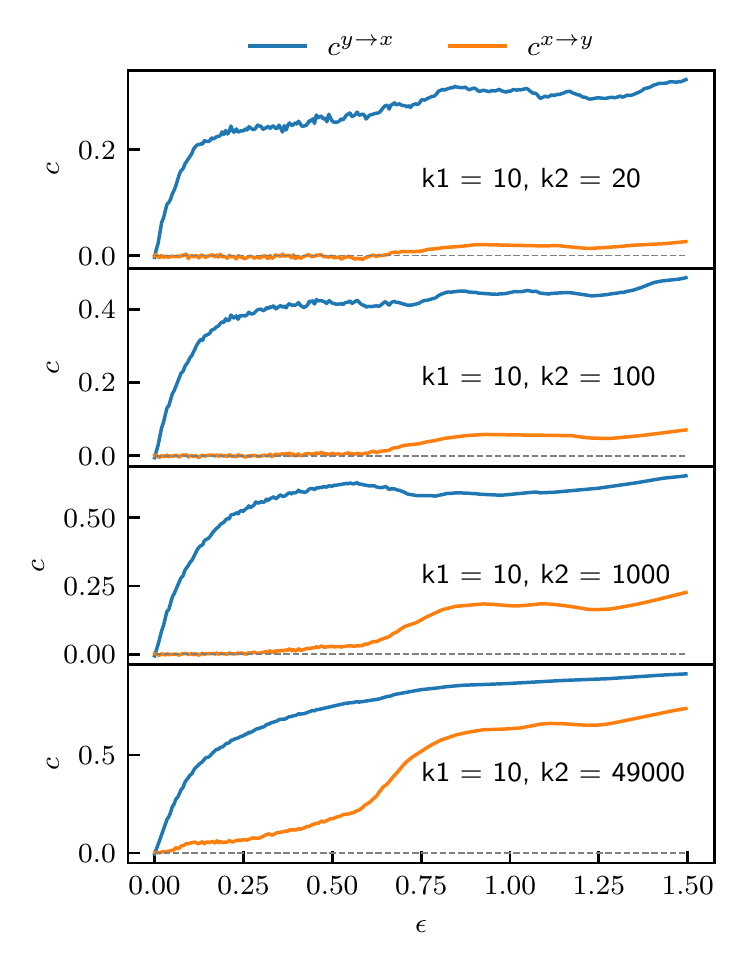}
\caption{The dependence of the coupling indices $c$ on the coupling parameter $\epsilon$ for different values of the parameter $k_2$ and fixed $k_1=10$.}
\label{k2dependence}
\end{figure}

\begin{acknowledgments}
The authors acknowledge the project J3-4525 and the research core funding No. P2-0001 that were financially supported by the Slovenian Research Agency. 
\end{acknowledgments}

\section*{Author declarations}
The experiment from Section \ref{sectiobio} was approved by the Ethical Committee of the University Medical Centre Ljubljana, Slovenia.
All subjects provided written informed consent.

\section*{Data availability}
The data that support the analysis of this article have been generated by the authors and can be fully reproduced from the repository \url{https://repo.ijs.si/mbresar/cd-vec} \cite{repo}.


\nocite{*}
\bibliography{detcoupling}

\end{document}